\documentclass[two column,rsi,amsmath,amssymb,superscriptaddress,prl,citeautoscript] {revtex4-1}
\setcitestyle{super}
\usepackage{color,graphics}
\usepackage{graphicx}
\usepackage[sort&compress]{natbib}
\usepackage{hhline}
\usepackage{soul}
\usepackage{xcolor}
\usepackage{array}
\usepackage{bm}

\usepackage[normalem]{ulem}
\usepackage{hyperref}

\usepackage[thinlines]{easytable}
\begin{document}
\setcounter{totalnumber}{3}
\renewcommand{\thetable}{\arabic{table}}
\newcolumntype{P}[1]{>{\centering\arraybackslash}p{#1}}

\title{Biexcitons are bound in CsPbBr$_3$ Perovskite Nanocrystals}

 \author{Yoonjae Park}
 \affiliation{Department of Chemistry, University of California, Berkeley, California 94720, USA}
 \affiliation{Materials Science Division, Lawrence Berkeley National Laboratory, Berkeley, California 94720, USA}

\author{David T. Limmer}
 \email{dlimmer@berkeley.edu}
 \affiliation{Department of Chemistry, University of California, Berkeley, California 94720, USA}
\affiliation{Materials Science Division, Lawrence Berkeley National Laboratory, Berkeley, California 94720, USA}
\affiliation{Chemical Science Division, Lawrence Berkeley National Laboratory, Berkeley, California 94720, USA}
\affiliation{Kavli Energy NanoScience Institute, Berkeley, California 94720, USA}

\date{\today}
\vspace{0mm}

\begin{abstract}
We study the energetics of quasi-particle excitations in CsPbBr$_3$ perovskite nanocrystals using path integral molecular dynamics simulations. Employing detailed molecular models, we elucidate the interplay of anharmonic lattice degrees of freedom, dielectric confinement, and electronic correlation on 
exciton and biexciton binding energies over a range of nanocrystal sizes. We find generally good agreement with some experimental observations on binding energies, and additionally explain the observed size dependent Stokes shift. The explicit model calculations are compared to simplified approximations to rationalize the lattice contributions to binding. We find that polaron formation significantly reduces exciton binding energies, whereas these effects are negligible for biexciton interactions. While experimentally, the binding energy of biexcitons is uncertain, based on our study we conclude that biexcitons are bound in CsPbBr$_3$.  
\end{abstract}

\maketitle


Lead halide perovskite nanocrystals are currently the subject of significant interest due to their exceptionally high photoluminescence quantum yields, which make them ideal materials for light emission, lasing and photodetection \cite{10.1039/c7tc05658c, 10.1039/d2cp02826c, 10.1002/adma.202208354, 10.1088/2515-7647/abedd0, 10.1021/acs.jpclett.9b03282}. 
The electronic properties of the lead halides depend strongly on the coupling between charges and their surrounding soft, polar lattices.\cite{joshi2019dynamic,egger2018remains,limmer2020photoinduced,egger2016hybrid,berry2015hybrid,schilcher2021significance}
This coupling has been implicated in a number of novel phenomena including photo-induced phase transitions, long radiative recombination rates, and anomalous temperature dependent mobilities\cite{bischak2017origin, mayers2018lattice, stranks2013electron, park2022,martin2023multiple}. In nanocrystals, optical properties are largely determined by the behavior of exciton complexes \cite{10.1021/acs.jpclett.9b00524}. Recently, Dana et al. \cite{dana2021unusually} proposed a potential anti-binding of biexcitons in perovskite nanocrystals, implicating the potential role of the lattice in mediating this interaction. Here, we study the energetics of quasiparticle excitations including excitons and biexcitons with an explicit description of the lattice, over a range of perovskite nanocrystal sizes. We find that while polaron formation weakens the exciton binding energy, biexciton energetics are largely unaffected, leading to an expectation that they are bound in nanocrystals and in bulk.

Excitons and biexcitons are both characterized by a binding energy, the energy required to dissociate the pair of quasiparticles--free charges or excitons. While there is reasonable consensus on the bulk exciton binding energy\cite{10.1021/nl5048779,10.1016/j.jcis.2018.12.105, 10.1063/1.5128016, 10.1021/acs.jpclett.7b00017, 10.1021/acs.jpclett.5b01252, 10.1021/acsomega.0c05414}, its dependence on nanocrystal size is less certain. Further, a number of biexciton binding energies have been reported experimentally for lead halide perovskites, but their values span a large range, and even disagree in sign\cite{10.1002/adma.202208354,10.1039/d2cp02826c, dana2021unusually, 10.1021/acsenergylett.9b02041, 10.1021/acs.nanolett.3c00793, 10.1021/acs.nanolett.5b05077, 10.1021/acs.nanolett.1c02122, 10.1021/acs.jpclett.8b01029, 10.1021/acs.jpcc.7b00762, 10.1021/acsnano.6b03908, dana2021unusually, 10.1021/acsenergylett.9b02041}. Experimental measurements are hampered by the nanocrystal polydispersity, spectral drift, and thermal broadening of spectral lines \cite{10.1021/acs.nanolett.1c01291, 10.1021/acs.nanolett.9b02856, 10.1103/physrevlett.111.177401}. 
These ambiguities could be clarified theoretically, however there are few suitable approaches available.
Unlike traditional semiconductors where structural fluctuations can be ignored or described within a harmonic approximation, the perovskite lattice structure with its anharmonic tilting and rocking motions result in significant renormalization of electronic properties.\cite{pollmann1977effective,haken1956quantum,parkjcp2022}  Given that incorporating the effects from the lattice is important in understanding excitonic properties, there have been some attempts to include them into a theoretical description. For example, extensions to GW and Bethe-Salpeter equations have been developed to include phonon effects on excitonic properties perturbatively.\cite{neatonprl,multiphonon, giustinoPRB, giustinopolaron,rohlfing2000electron, albrecht1998excitonic, hedin1965new} At the same time, quasiparticle excitations requires a balanced description of electron correlation, making \emph{ab initio} models difficult to apply in nanocrystals where the number of atoms are large. Tight binding and pseudo-potential models have been developed to study excitonic structure, but these have not yet been unified with approaches to describe electron-phonon effects.\cite{cho2019optical,biffi2023excitons, cho2021simulations,danielarxiv} 

Here, we use quasiparticle path integral molecular dynamics\cite{parrinello1984study} with an explicit atomistic description of the lattice, which allows us to include all orders of anharmonicity from the lattice, and treat electron correlation exactly. This approach has been previously successful at describing the excitonic properties of bulk systems,\cite{park2022,parkjcp2022} and the lattice model we employ has been used extensively to describe both vibrational and nonlinear properties of a variety of lead-halide systems.\cite{danielarxiv, 10.1016/j.matt.2020.07.015, gao2023direct, quan2021vibrational}
In this study, we consider a system of a single biexciton interacting with CsPbBr$_3$ cubic perovskite nanocrystals. The model Hamiltonian consists of three pieces, $\mathcal{H} = \mathcal{H}_{\mathrm{el}} + \mathcal{H}_{\mathrm{lat}} + \mathcal{H}_{\mathrm{int}}$. Within an effective mass approximation, valid because of the highly dispersive bands,\cite{sajedi2022there} the electronic Hamiltonoan $\mathcal{H}_{\mathrm{el}}$ is defined by kinetic energies and Coulomb interactions between electrons and holes, 
\begin{equation}
{\mathcal{H}}_{\mathrm{el}} 
= \sum_{i} \frac{\hat{\mathbf{p}}_i^2}{2m_i} 
+ \sum_{i \ne j} \frac{q_i q_j}{4\pi \varepsilon_{\infty} |\hat{\mathbf{x}}_{i} - \hat{\mathbf{x}}_{j}|}
\label{biex_Hel}
\end{equation}
where the subscripts $i,j \in \{e_1, e_2, h_1, h_2 \}$ indicate two electrons and two holes respectively, $\hat {\mathbf{p}}$ and $\hat {\mathbf{x}}$ are momentum and position operators, $m_i$ is the band mass  set to 0.22\,$m_0$ and 0.24\,$m_0$ for electrons and holes in terms of the bare electron mass $m_0$,
 $\varepsilon_{\infty}=4.3$ is the optical dielectric constant in the unit of vacuum permittivity $\varepsilon_0$, and $q$ is the charge of quantum particle which is $-e$ for electrons and $+e$ for holes.\cite{sciadvzhu} We consider a singlet biexciton, so electrons and holes are distinguishable particles.\cite{ceperley1995path} 

For the lattice, we model the CsPbBr$_3$ nanocrystals explicitly using a previously validated \emph{ab initio} derived forcefield.\cite{10.1016/j.matt.2020.07.015,danielarxiv} Its Hamiltonian is given by
\begin{equation}
{\mathcal{H}}_{\mathrm{lat}}
= \sum_{i=1}^N \frac{\mathbf{p}_i^2}{2m_i} 
+ U_{\mathrm{lat}}(\mathbf{x}_{\mathrm{lat}})
\label{biex_Hlat}
\end{equation}
where $\mathbf{x}_{\mathrm{lat}} = \{\mathbf{x}_1, \mathbf{x}_2, \dots, \mathbf{x}_N \}$ are the positions of the $N$ atoms in the lattice, $\mathbf{p}_i$ and $m_i$ are the momentum and mass for $i^{th}$ atom. The atomistic forcefield \cite{10.1016/j.matt.2020.07.015}, $U_{\mathrm{lat}}$, is the sum of pairwise interactions with distance $x_{ij}=|\mathbf{x}_i-\mathbf{x}_j|$ consisting of a Coulomb potential and Lennard-Jones potential 
\begin{equation}
U_{\mathrm{lat}} = \sum_{i,j=1}^N \frac{q_i q_j}{4 \pi \varepsilon_0 x_{ij}} + 4\varepsilon_{ij} \left[ \left( \frac{\sigma_{ij}}{x_{ij}} \right)^{12} - \left( \frac{\sigma_{ij}}{x_{ij}} \right)^{6} \right]
\end{equation}
where the parameters $q_i$, $\varepsilon_{ij}$ and $\sigma_{ij}$ are summarized in Table S1. The interaction between the quasiparticles and lattice is written as the sum of pseudopotentials 
between each quantum particle $i$ and the lattice particle $j$, 
\begin{equation}
\mathcal{H}_{\mathrm{int}} = \sum_{i} \sum_{j=1}^N \frac{q_i q_j}{4\pi \varepsilon_0 \sqrt{ r_{\mathrm{cut}}^2 +|\hat{\mathbf{x}}_{i}-\hat{\mathbf{x}}_j|^2} }
\end{equation}
 where the cutoff distances $r_{\mathrm{cut}}$ are chosen using the atomic radii of each atoms \cite{park2022, parrinello1984study, schnitker1987electron, kuharski1988molecular}. 
 
The quasiparticle interactions are screened by the optical dielectric constant of the nanocrystal, which changes discontinuously at the boundary between the perovskite and surrounding solution. 
To account for the dielectric discontinuity, we use a multipole expansion,\cite{wang1994dielectric} resulting in an effective potential for each charge,  
 \begin{equation}
 U_{\mathrm{wall}} = \sum_{i,k} \frac{q_i^2}{8 \pi \varepsilon_0 | \hat{\mathbf{x}}_{i} - \mathbf{x}_{\mathrm{wall},k}|} \cdot \frac{\varepsilon_{\infty}-\varepsilon_{0}}{\varepsilon_{\infty}+\varepsilon_{0}}
 \end{equation} 
where $i \in \{e_1, e_2, h_1, h_2 \}$, $k \in \{ \pm x, \pm y, \pm z \}$, and $\mathbf{x}_{\mathrm{wall},k} = \pm L/2$ is the position of the wall with $L$ as the edge length of the nanocrystal. This external potential results in a dielectric confinement. Additional details on the atomistic force field, pseudopotentials, and wall potentials are described in the Supporting information (SI). 

\begin {figure}
\centering\includegraphics [width=8.5cm] {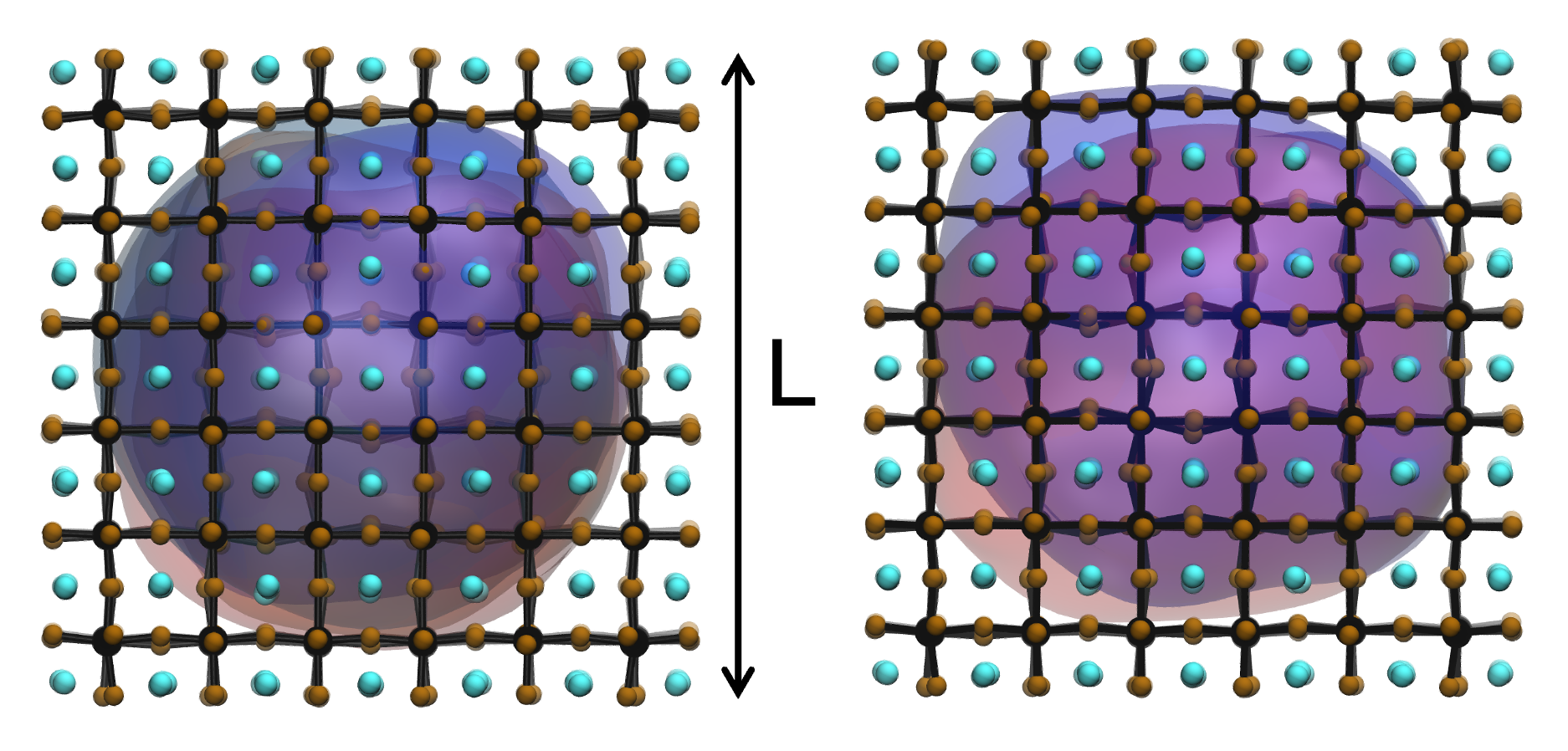}
\caption{Simulation snapshot of biexciton (left) and exciton (right) interacting with CsPbBr$_3$ perovskite nanocrystal where $L=3.56$ nm represents the edge length of the nanocrystal. }
\label{vmd}
\end{figure}

For electrons and holes, we use imaginary time path integrals,\cite{feynman1, feynman2} while the heavy atoms of the lattice are treated classically. The resultant partition function $\mathcal{Z}$ can be written as
\begin{equation}
\mathcal{Z} 
= \int \mathcal{D} [ \mathbf{x}_{eh}(\tau), \, \mathbf{x}_{\mathrm{lat}}] \, e^{-\int_{\tau = 0}^{\beta \hbar} \mathcal{H}[ \mathbf{x}_{eh}(\tau), \, \mathbf{x}_{\mathrm{lat}}]/\hbar} 
\label{biex_pfn}
\end{equation}
where $\mathbf{x}_{eh}(\tau) = \{\mathbf{x}_{e_1}(\tau), \mathbf{x}_{e_2}(\tau), \mathbf{x}_{h_1}(\tau), \mathbf{x}_{h_2}(\tau) \}$  where $\mathbf{x}_{i}(\tau)$ is the position of quantum particle $i$ at imaginary time $\tau$, $\hbar$ is the Planck's constant, and $\beta^{-1} = k_{\mathrm{B}}T$ with $k_{\mathrm{B}}$ and $T$ as the Boltzmann constant and temperature. Discretizing the path action renders each quantum particle isomorphic to a ring polymer, where neighboring timeslices are connected by harmonic springs\cite{habershon2013ring, ceperley1995path}. We use molecular dynamics simulations with a second order discretization of the path integral to compute expectation values of this system, for which each quant particle is represented by 1000 timeslices. Representative simulation snapshots of both the biexciton and exciton are shown in Fig.~\ref{vmd}. Throughout we consider nano-crystalline cubes, where $L$ is the edge length.

For comparison, we also consider two approximate models that incorporate harmonic and static lattice effects. 
To incorporate effects from harmonic phonons, we adopt a \textit{dynamic} screening model described previously\cite{park2022, parkjcp2022}, which is a model for charges that are coupled linearly with the polarization field generated by a collection of harmonic modes. In this model, lattice variables are analytically integrated out, resulting in an imaginary time influence functional that can be studied straightforwardly numerically. We parameterize the influence functional with a Frohlich coupling for the electrons and holes as $\alpha_e=2.65$ and {$\alpha_h=2.76$} and an optical phonon mode with energy {$\hbar \omega=16.8$ m\emph{e}V}. \cite{sciadvzhu}
We also compare a \textit{static} lattice, where only the electronic Hamiltonian and confinement effects are considered. 
All simulations are done in LAMMPS \cite{lammps} and the details of the discretized Hamiltonian can be found in the SI. 

With an explicit lattice, the exciton binding energy is 
$$
\Delta_{{X}} =\lim_{T\rightarrow 0}  \langle E \rangle_{\mathrm{e}} + \langle E \rangle_{\mathrm{h}} - \langle E \rangle_{\mathrm{ex}} -  \langle U_{\mathrm{lat}} \rangle_{\mathrm{ex}} \, ,
$$ while the biexciton binding energy is 
$$
\Delta_{{XX}} =  \lim_{T\rightarrow 0} 2 \langle E \rangle_{\mathrm{ex}} - \langle E \rangle_{\mathrm{biex}} - \langle U_{\mathrm{lat}} \rangle_{\mathrm{ex}}  \, ,
$$
where the subscripts  $\mathrm{biex}$ and $\mathrm{ex}$ refers to a simulation of a biexciton (two electrons and two holes) and exciton (electron and hole), while $\mathrm{e} / \mathrm{h}$ indicates a simulation with only electron/hole interacting with the surrounding lattice. {Simulations are performed at 50K, which is low enough to extract the ground state energy.} The binding energies is computed from the average energy \cite{binde},
\begin{equation}
\langle E \rangle = -\frac{\partial }{\partial \beta}
\ln \mathcal{Z} [\mathbf{x}_{eh}(\tau), \mathbf{x}_{\mathrm{lat}}(\tau)] 
\label{biex_avgE}
\end{equation}
where $\mathbf{x}_{eh} = \{\mathbf{x}_{e_1}, \mathbf{x}_{h_1} \}$ or $\mathbf{x}_{eh} = \{\mathbf{x}_{e_1}, \mathbf{x}_{e_2}, \mathbf{x}_{h_1}, \mathbf{x}_{h_2} \}$ for exciton and biexciton, respectively. The derivative above produces two terms, an average kinetic energy and an average potential energy, where we use a virial estimator\cite{virial} to efficiently estimate the kinetic energy. For the evaluation of binding energies from the simulations with only quasiparticles, in the static or dynamic approximation, the same definitions are used without the lattice relevant terms. Within the static approximation, the binding energies are determined by only $\mathcal{H}_{e}$ and the confining potential. In order to efficiently extract the binding energies with dynamic approximation, we use a thermodynamic perturbative theory approach (see SI).


\begin {figure}
\centering\includegraphics [width=8.9cm] {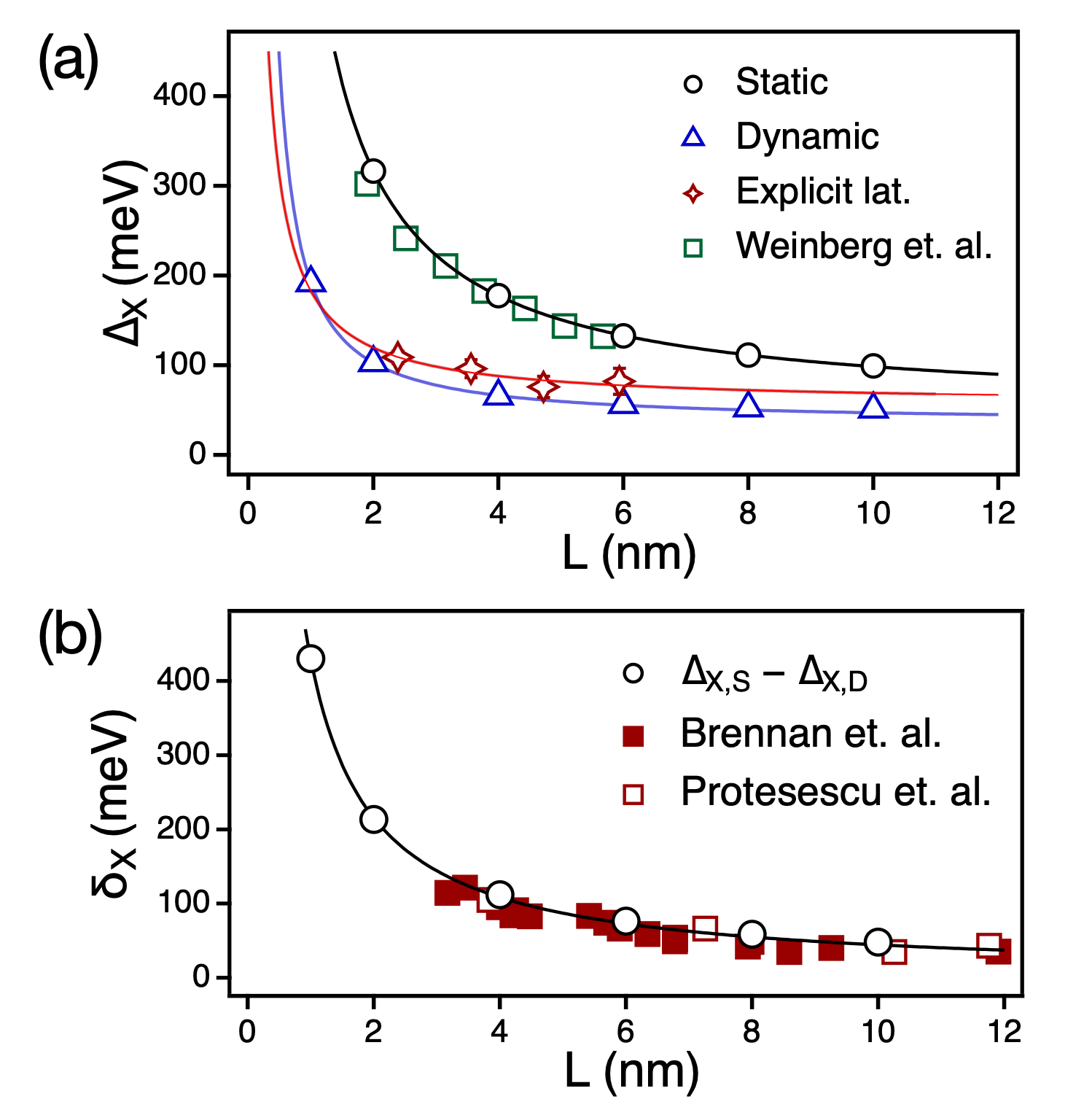}
\caption{Energetics of exciton formation. (a) Exciton binding energy under static ($\Delta_{X,S}$ black circles) dynamic ($\Delta_{X,D}$ blue triangles) and explicit lattice (red stars) models. Comparison to Weinberg et al\cite{danielarxiv} is shown in {green} squares. 
(b) The difference in the exciton binding energies from dynamic and static screenings, which is defined as $\delta_{\mathrm{X}} = \Delta_{\mathrm{X,S}} - \Delta_{\mathrm{X,D}}$ compared to Stokes shift measurements from Brennan et al\cite{10.1021/acs.jpclett.0c01407} and Protesescu et. al\cite{protesescu2015nanocrystals}. Solid lines are fits to $\Delta_{X}=\Delta_{X}^{\circ} (1 + \ell_1/L + \ell_2^2/L^2)$ and $\delta_X=\delta_{X}^{\circ}(1+\ell'/L)$. }
\label{energy}
\end{figure}

Figure~\ref{energy} (a) shows the exciton binding energy $\Delta_{\mathrm{X}}$ computed from our molecular model, in addition to those approximated from with the static or dynamics models for the nanocubes considered, ranging from $L=$ 2.4\,nm to 6\,nm. Given the reduced mass of the exciton, $\mu=0.11$ and optical dielectric constant, the Bohr radius of the exciton is $R_X=2.07$nm. Thus for the range of nanocrystals considered, we are in a moderate to strong confinement regime. As a consequence, all binding energy estimates exhibit a strong $L$ dependence, which we model in Fig.~\ref{energy}(a) as $\Delta_{X}=\Delta_{X}^{\circ} (1 + \ell_1/L + \ell_2^2/L^2)$ where $\Delta_{X}^{\circ}$ is the bulk binding energy and $\ell_1$ and $\ell_2$ are fit parameters.\cite{biexfitting, 10.1103/physrevlett.64.1805}  

Over the full range of nanocrystal sizes considered, the static approximation yields a binding energy much larger than either the explicit model calculation or the dynamic approximation. 
{The value from the static approximation with the largest nanocrystal gives a reasonable agreement with the expectation from a Wannier-Mott model $\Delta_{X,\textrm{WM}} = R_H \mu/\varepsilon_{\infty}^2 = $ 84.5 m\emph{e}V, where $R_H$ is a Rydberg energy}. The values at finite $L$ agree well with recent pseudo-potential based calculations of CsPbI$_3$, up to a shift of the bulk binding energy by {20 m\emph{e}V} to account for the change in halide.\cite{danielarxiv}

The explicit model calculations are well reproduced by the dynamic approximation. The suppression of the binding energy due to lattice effects could arise because of two distinct mechanisms. First the lattice could screen the electron-hole interactions, weakening them. Second the lattice can lower the self-energies of the free charges, bringing them closer to the exciton self-energy, effectively reducing the binding energy. In agreement with previous calculations on bulk MAPbI$_3$,\cite{parkjcp2022} we find the latter effect is more prominent.
This effect is outside of first order perturbation, which over estimates the decrease in the self-energy of the free charges which we find is -30 m\emph{e}V for both charges, and ignores the decrease in self-energy of the polaron-exciton, which is -12 m\emph{e}V.
With the explicit perovskite lattice, the extrapolated value to the large nanocrystal size limit is in good agreement with bulk CsPbBr$_3$ exciton binding energy of 40 meV, reflecting slight anharmonic weakening of the optical phonon\cite{10.1021/nl5048779,10.1016/j.jcis.2018.12.105, 10.1063/1.5128016, 10.1021/acs.jpclett.7b00017, 10.1021/acs.jpclett.5b01252, 10.1021/acsomega.0c05414}.

Shown in Fig.~\ref{energy} (b) is the difference between the dynamic and static screening models, $\delta_X = \Delta_{X,S}-\Delta_{X,D}$. This energy is due to polaron formation. The strong nanocrystal size dependence reflects the increasing confinement of the charges, which lower their self-energy by increasing their localization, as the energy of a charge in a dielectric will decrease like $1/L$, which fits $\delta_X$ well. We find the polaron formation energy agrees well with size-dependent Stokes shift measurements made from the difference between absorption and emission spectra.\cite{10.1038/s41467-019-09057-5} Results from two different experiments on CsPbBr$_3$ nanocrystals \cite{protesescu2015nanocrystals, 10.1021/jacs.7b05683}, compare quantitatively well with our computed $\delta_X$, increasing from 10 m\emph{e}V to over 450 m\emph{e}V. Previously, a lattice origin of the Stokes shift had been disregarded due to the relatively small polaron formation energy in bulk.\cite{10.1021/jacs.7b05683} However, we find the polaron formation energy can increase substantially in small nanocrystals. 

\begin {figure}
\centering\includegraphics [width=8.9cm] {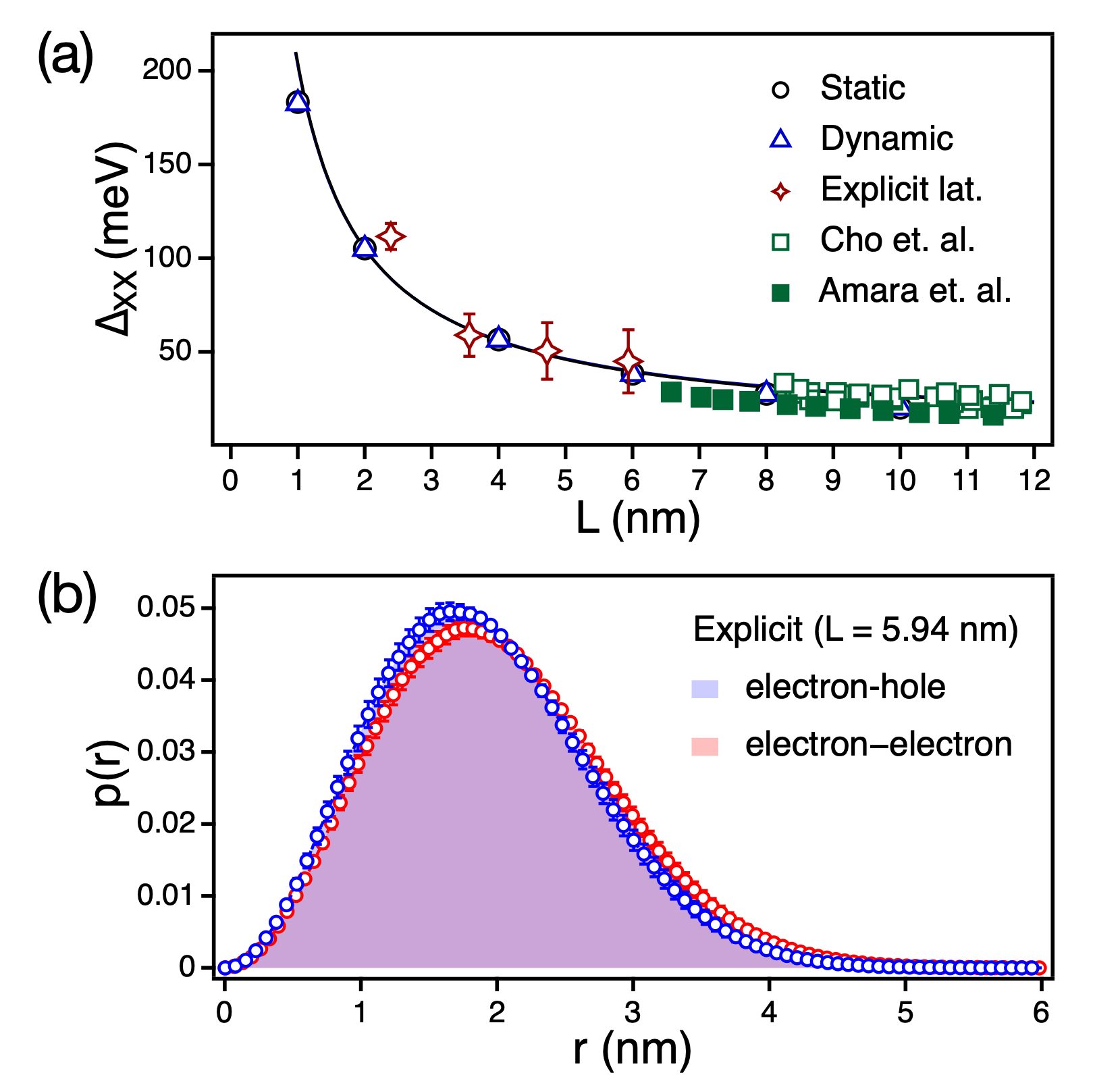}
\caption{Energetics and structure of biexcitons. (a) Biexciton binding energies for our explicit model (red stars) and its dynamical (blue squares) and static (black circles) approximations. Comparisons to Cho. et al \cite{10.1021/acs.nanolett.1c02122} and Amara et. al\cite{10.1021/acs.nanolett.3c00793} are shown in green squares. {The solid line is a fit to $\Delta_{XX}=\Delta_{XX}^{\circ} (1 + \tilde{\ell_1}/L + \tilde{\ell_2}^2/L^2)$}. (b) Electron-hole (blue) and electron-electron (red) distributions of biexciton from the simulations with explicit lattice.  }
\label{pofr}
\end{figure}

With the excitonic energetics understood, we now turn to the binding of biexcitons. The biexciton binding energy $\Delta_{XX}$ for the explicit model, as well as the static and dynamic approximations, are shown in Fig.~\ref{pofr}(a). All three models agree quantitatively, indicating that contrary from expectations from the exciton calculations, lattice effects are unimportant for biexciton binding. We rationalize this by noting that the primary contribution from the lattice for the exciton was the stabilization of the free charges due to the polaron formation, which does not contribute to the biexciton energy. In Fig.~\ref{pofr} (b) is the electron-hole and electron-electron radial probability distributions, $p(r)$, from the simulations with explicit lattice where electron-electron repulsion makes the average distance slightly larger than the average electron-hole distance. However this increase is small and the their size is largely determined by confinement rather than electron correlation. The similar masses for the electron and hole, produce a very weak dipole in the exciton or quadrapole in the biexcton that is not large enough to generate a polarization response from the lattice. Further, the similar results in both binding energies from explicit lattice and dynamic models imply that the anharmonicity coming from the explicit perovskite lattice does not play a crucial role in determining the behavior of biexcitons, likely due to the relatively small amplitude distortion. However, $\Delta_{XX}$ is relatively large, 1/2-1/4 of $\Delta_X$, indicating the importance of electron correlation. 

As with $\Delta_X$, we find a strong system size dependence of $\Delta_{XX}$, increasing strongly with decreasing $L$. {We model this dependence as $\Delta_{XX}=\Delta_{XX}^{\circ}(1+\tilde{\ell}_1/L + \tilde{\ell_2}^2/L^2)$, with $\tilde{\ell}_1$ and $\tilde{\ell}_2$ employed as fitting parameters.} While experimentally values of the $\Delta_{XX}$ have been reported between -100 and 100 m\emph{e}V,\cite{10.1002/adma.202208354,10.1039/d2cp02826c, dana2021unusually, 10.1021/acsenergylett.9b02041, 10.1021/acs.nanolett.3c00793, 10.1021/acs.nanolett.5b05077, 10.1021/acs.nanolett.1c02122, 10.1021/acs.jpclett.8b01029, 10.1021/acs.jpcc.7b00762, 10.1021/acsnano.6b03908} recent size dependent measurements on CsPbBr$_3$ nanocubes from two different groups\cite{10.1021/acs.nanolett.3c00793,10.1021/acs.nanolett.1c02122} are shown in Fig.~\ref{pofr}(a) and agree well with our results. 

In summary, we employed path integral molecular dynamics simulations to study quasiparticle energetics in perovskite nanocrystals. Using atomic models and simple approximations, we are able to systematically explore the roles of the lattice on effects on exciton and biexciton binding energies. We found that the effects from the lattice are large in exciton binding energy. but negligible in biexciton interaction, attributed to the weak coupling between the lattice and the neutral exciton complex.  For all nanocrystals, regardless of the size and the type of screening employed, we found that biexcitons are bound, strongly suggestive that reports of anti-binding are misinterpreted. 

\emph{Acknowledgements.} This work was supported by the U.S. Department of Energy,
Office of Science, Office of Basic Energy Sciences, Materials Sciences and Engineering Division, under Contract No. DEAC02-05-CH11231 (Grant No. KC3103). Y.P. acknowledges the Kwanjeong Educational
Foundation. D.T.L. acknowledges the Alfred P. Sloan Foundation.

\vspace{1mm}

\bibliography{Bmain}

\begin{thebibliography}{70}%
\makeatletter
\providecommand \@ifxundefined [1]{%
 \@ifx{#1\undefined}
}%
\providecommand \@ifnum [1]{%
 \ifnum #1\expandafter \@firstoftwo
 \else \expandafter \@secondoftwo
 \fi
}%
\providecommand \@ifx [1]{%
 \ifx #1\expandafter \@firstoftwo
 \else \expandafter \@secondoftwo
 \fi
}%
\providecommand \natexlab [1]{#1}%
\providecommand \enquote  [1]{``#1''}%
\providecommand \bibnamefont  [1]{#1}%
\providecommand \bibfnamefont [1]{#1}%
\providecommand \citenamefont [1]{#1}%
\providecommand \href@noop [0]{\@secondoftwo}%
\providecommand \href [0]{\begingroup \@sanitize@url \@href}%
\providecommand \@href[1]{\@@startlink{#1}\@@href}%
\providecommand \@@href[1]{\endgroup#1\@@endlink}%
\providecommand \@sanitize@url [0]{\catcode `\\12\catcode `\$12\catcode
  `\&12\catcode `\#12\catcode `\^12\catcode `\_12\catcode `\%12\relax}%
\providecommand \@@startlink[1]{}%
\providecommand \@@endlink[0]{}%
\providecommand \url  [0]{\begingroup\@sanitize@url \@url }%
\providecommand \@url [1]{\endgroup\@href {#1}{\urlprefix }}%
\providecommand \urlprefix  [0]{URL }%
\providecommand \Eprint [0]{\href }%
\providecommand \doibase [0]{http://dx.doi.org/}%
\providecommand \selectlanguage [0]{\@gobble}%
\providecommand \bibinfo  [0]{\@secondoftwo}%
\providecommand \bibfield  [0]{\@secondoftwo}%
\providecommand \translation [1]{[#1]}%
\providecommand \BibitemOpen [0]{}%
\providecommand \bibitemStop [0]{}%
\providecommand \bibitemNoStop [0]{.\EOS\space}%
\providecommand \EOS [0]{\spacefactor3000\relax}%
\providecommand \BibitemShut  [1]{\csname bibitem#1\endcsname}%
\let\auto@bib@innerbib\@empty
\bibitem [{\citenamefont {Hong}\ \emph {et~al.}(2018)\citenamefont {Hong},
  \citenamefont {Le}, \citenamefont {Kim},\ and\ \citenamefont
  {Jang}}]{10.1039/c7tc05658c}%
  \BibitemOpen
  \bibfield  {author} {\bibinfo {author} {\bibfnamefont {K.}~\bibnamefont
  {Hong}}, \bibinfo {author} {\bibfnamefont {Q.~V.}\ \bibnamefont {Le}},
  \bibinfo {author} {\bibfnamefont {S.~Y.}\ \bibnamefont {Kim}}, \ and\
  \bibinfo {author} {\bibfnamefont {H.~W.}\ \bibnamefont {Jang}},\ }\href
  {\doibase 10.1039/c7tc05658c} {\bibfield  {journal} {\bibinfo  {journal}
  {Journal of Materials Chemistry C}\ }\textbf {\bibinfo {volume} {6}},\
  \bibinfo {pages} {2189} (\bibinfo {year} {2018})}\BibitemShut {NoStop}%
\bibitem [{\citenamefont {Yumoto}\ and\ \citenamefont
  {Kanemitsu}(2022)}]{10.1039/d2cp02826c}%
  \BibitemOpen
  \bibfield  {author} {\bibinfo {author} {\bibfnamefont {G.}~\bibnamefont
  {Yumoto}}\ and\ \bibinfo {author} {\bibfnamefont {Y.}~\bibnamefont
  {Kanemitsu}},\ }\href {\doibase 10.1039/d2cp02826c} {\bibfield  {journal}
  {\bibinfo  {journal} {Physical Chemistry Chemical Physics}\ }\textbf
  {\bibinfo {volume} {24}},\ \bibinfo {pages} {22405} (\bibinfo {year}
  {2022})}\BibitemShut {NoStop}%
\bibitem [{\citenamefont {Zhu}\ \emph {et~al.}(2023)\citenamefont {Zhu},
  \citenamefont {Nguyen}, \citenamefont {Boehme}, \citenamefont {Moskalenko},
  \citenamefont {Dirin}, \citenamefont {Bodnarchuk}, \citenamefont {Katan},
  \citenamefont {Even}, \citenamefont {Rainò},\ and\ \citenamefont
  {Kovalenko}}]{10.1002/adma.202208354}%
  \BibitemOpen
  \bibfield  {author} {\bibinfo {author} {\bibfnamefont {C.}~\bibnamefont
  {Zhu}}, \bibinfo {author} {\bibfnamefont {T.}~\bibnamefont {Nguyen}},
  \bibinfo {author} {\bibfnamefont {S.~C.}\ \bibnamefont {Boehme}}, \bibinfo
  {author} {\bibfnamefont {A.}~\bibnamefont {Moskalenko}}, \bibinfo {author}
  {\bibfnamefont {D.~N.}\ \bibnamefont {Dirin}}, \bibinfo {author}
  {\bibfnamefont {M.~I.}\ \bibnamefont {Bodnarchuk}}, \bibinfo {author}
  {\bibfnamefont {C.}~\bibnamefont {Katan}}, \bibinfo {author} {\bibfnamefont
  {J.}~\bibnamefont {Even}}, \bibinfo {author} {\bibfnamefont {G.}~\bibnamefont
  {Rainò}}, \ and\ \bibinfo {author} {\bibfnamefont {M.~V.}\ \bibnamefont
  {Kovalenko}},\ }\href {\doibase 10.1002/adma.202208354} {\bibfield  {journal}
  {\bibinfo  {journal} {Advanced Materials}\ }\textbf {\bibinfo {volume}
  {35}},\ \bibinfo {pages} {2208354} (\bibinfo {year} {2023})}\BibitemShut
  {NoStop}%
\bibitem [{\citenamefont {Ahumada-Lazo}\ \emph {et~al.}(2021)\citenamefont
  {Ahumada-Lazo}, \citenamefont {Saran}, \citenamefont {Woolland},
  \citenamefont {Jia}, \citenamefont {Kyriazi}, \citenamefont {Kanaras},
  \citenamefont {Binks},\ and\ \citenamefont
  {Curry}}]{10.1088/2515-7647/abedd0}%
  \BibitemOpen
  \bibfield  {author} {\bibinfo {author} {\bibfnamefont {R.}~\bibnamefont
  {Ahumada-Lazo}}, \bibinfo {author} {\bibfnamefont {R.}~\bibnamefont {Saran}},
  \bibinfo {author} {\bibfnamefont {O.}~\bibnamefont {Woolland}}, \bibinfo
  {author} {\bibfnamefont {Y.}~\bibnamefont {Jia}}, \bibinfo {author}
  {\bibfnamefont {M.-E.}\ \bibnamefont {Kyriazi}}, \bibinfo {author}
  {\bibfnamefont {A.~G.}\ \bibnamefont {Kanaras}}, \bibinfo {author}
  {\bibfnamefont {D.}~\bibnamefont {Binks}}, \ and\ \bibinfo {author}
  {\bibfnamefont {R.~J.}\ \bibnamefont {Curry}},\ }\href {\doibase
  10.1088/2515-7647/abedd0} {\bibfield  {journal} {\bibinfo  {journal} {Journal
  of Physics: Photonics}\ }\textbf {\bibinfo {volume} {3}},\ \bibinfo {pages}
  {021002} (\bibinfo {year} {2021})}\BibitemShut {NoStop}%
\bibitem [{\citenamefont {Vale}\ \emph {et~al.}(2020)\citenamefont {Vale},
  \citenamefont {Socie}, \citenamefont {Burgos-Caminal}, \citenamefont
  {Bettini}, \citenamefont {Schiavon},\ and\ \citenamefont
  {Moser}}]{10.1021/acs.jpclett.9b03282}%
  \BibitemOpen
  \bibfield  {author} {\bibinfo {author} {\bibfnamefont {B.~R.~C.}\
  \bibnamefont {Vale}}, \bibinfo {author} {\bibfnamefont {E.}~\bibnamefont
  {Socie}}, \bibinfo {author} {\bibfnamefont {A.}~\bibnamefont
  {Burgos-Caminal}}, \bibinfo {author} {\bibfnamefont {J.}~\bibnamefont
  {Bettini}}, \bibinfo {author} {\bibfnamefont {M.~A.}\ \bibnamefont
  {Schiavon}}, \ and\ \bibinfo {author} {\bibfnamefont {J.-E.}\ \bibnamefont
  {Moser}},\ }\href {\doibase 10.1021/acs.jpclett.9b03282} {\bibfield
  {journal} {\bibinfo  {journal} {The Journal of Physical Chemistry Letters}\
  }\textbf {\bibinfo {volume} {11}},\ \bibinfo {pages} {387} (\bibinfo {year}
  {2020})}\BibitemShut {NoStop}%
\bibitem [{\citenamefont {Joshi}\ \emph {et~al.}(2019)\citenamefont {Joshi},
  \citenamefont {Maehrlein},\ and\ \citenamefont {Zhu}}]{joshi2019dynamic}%
  \BibitemOpen
  \bibfield  {author} {\bibinfo {author} {\bibfnamefont {P.~P.}\ \bibnamefont
  {Joshi}}, \bibinfo {author} {\bibfnamefont {S.~F.}\ \bibnamefont
  {Maehrlein}}, \ and\ \bibinfo {author} {\bibfnamefont {X.}~\bibnamefont
  {Zhu}},\ }\href@noop {} {\bibfield  {journal} {\bibinfo  {journal} {Advanced
  Materials}\ }\textbf {\bibinfo {volume} {31}},\ \bibinfo {pages} {1803054}
  (\bibinfo {year} {2019})}\BibitemShut {NoStop}%
\bibitem [{\citenamefont {Egger}\ \emph {et~al.}(2018)\citenamefont {Egger},
  \citenamefont {Bera}, \citenamefont {Cahen}, \citenamefont {Hodes},
  \citenamefont {Kirchartz}, \citenamefont {Kronik}, \citenamefont {Lovrincic},
  \citenamefont {Rappe}, \citenamefont {Reichman},\ and\ \citenamefont
  {Yaffe}}]{egger2018remains}%
  \BibitemOpen
  \bibfield  {author} {\bibinfo {author} {\bibfnamefont {D.~A.}\ \bibnamefont
  {Egger}}, \bibinfo {author} {\bibfnamefont {A.}~\bibnamefont {Bera}},
  \bibinfo {author} {\bibfnamefont {D.}~\bibnamefont {Cahen}}, \bibinfo
  {author} {\bibfnamefont {G.}~\bibnamefont {Hodes}}, \bibinfo {author}
  {\bibfnamefont {T.}~\bibnamefont {Kirchartz}}, \bibinfo {author}
  {\bibfnamefont {L.}~\bibnamefont {Kronik}}, \bibinfo {author} {\bibfnamefont
  {R.}~\bibnamefont {Lovrincic}}, \bibinfo {author} {\bibfnamefont {A.~M.}\
  \bibnamefont {Rappe}}, \bibinfo {author} {\bibfnamefont {D.~R.}\ \bibnamefont
  {Reichman}}, \ and\ \bibinfo {author} {\bibfnamefont {O.}~\bibnamefont
  {Yaffe}},\ }\href@noop {} {\bibfield  {journal} {\bibinfo  {journal}
  {Advanced Materials}\ }\textbf {\bibinfo {volume} {30}},\ \bibinfo {pages}
  {1800691} (\bibinfo {year} {2018})}\BibitemShut {NoStop}%
\bibitem [{\citenamefont {Limmer}\ and\ \citenamefont
  {Ginsberg}(2020)}]{limmer2020photoinduced}%
  \BibitemOpen
  \bibfield  {author} {\bibinfo {author} {\bibfnamefont {D.~T.}\ \bibnamefont
  {Limmer}}\ and\ \bibinfo {author} {\bibfnamefont {N.~S.}\ \bibnamefont
  {Ginsberg}},\ }\href@noop {} {\bibfield  {journal} {\bibinfo  {journal} {The
  Journal of chemical physics}\ }\textbf {\bibinfo {volume} {152}} (\bibinfo
  {year} {2020})}\BibitemShut {NoStop}%
\bibitem [{\citenamefont {Egger}\ \emph {et~al.}(2016)\citenamefont {Egger},
  \citenamefont {Rappe},\ and\ \citenamefont {Kronik}}]{egger2016hybrid}%
  \BibitemOpen
  \bibfield  {author} {\bibinfo {author} {\bibfnamefont {D.~A.}\ \bibnamefont
  {Egger}}, \bibinfo {author} {\bibfnamefont {A.~M.}\ \bibnamefont {Rappe}}, \
  and\ \bibinfo {author} {\bibfnamefont {L.}~\bibnamefont {Kronik}},\
  }\href@noop {} {\bibfield  {journal} {\bibinfo  {journal} {Accounts of
  chemical research}\ }\textbf {\bibinfo {volume} {49}},\ \bibinfo {pages}
  {573} (\bibinfo {year} {2016})}\BibitemShut {NoStop}%
\bibitem [{\citenamefont {Berry}\ \emph {et~al.}(2015)\citenamefont {Berry},
  \citenamefont {Buonassisi}, \citenamefont {Egger}, \citenamefont {Hodes},
  \citenamefont {Kronik}, \citenamefont {Loo}, \citenamefont {Lubomirsky},
  \citenamefont {Marder}, \citenamefont {Mastai}, \citenamefont {Miller} \emph
  {et~al.}}]{berry2015hybrid}%
  \BibitemOpen
  \bibfield  {author} {\bibinfo {author} {\bibfnamefont {J.}~\bibnamefont
  {Berry}}, \bibinfo {author} {\bibfnamefont {T.}~\bibnamefont {Buonassisi}},
  \bibinfo {author} {\bibfnamefont {D.~A.}\ \bibnamefont {Egger}}, \bibinfo
  {author} {\bibfnamefont {G.}~\bibnamefont {Hodes}}, \bibinfo {author}
  {\bibfnamefont {L.}~\bibnamefont {Kronik}}, \bibinfo {author} {\bibfnamefont
  {Y.-L.}\ \bibnamefont {Loo}}, \bibinfo {author} {\bibfnamefont
  {I.}~\bibnamefont {Lubomirsky}}, \bibinfo {author} {\bibfnamefont {S.~R.}\
  \bibnamefont {Marder}}, \bibinfo {author} {\bibfnamefont {Y.}~\bibnamefont
  {Mastai}}, \bibinfo {author} {\bibfnamefont {J.~S.}\ \bibnamefont {Miller}},
  \emph {et~al.},\ }\href@noop {} {\bibfield  {journal} {\bibinfo  {journal}
  {Advanced Materials}\ }\textbf {\bibinfo {volume} {27}},\ \bibinfo {pages}
  {5102} (\bibinfo {year} {2015})}\BibitemShut {NoStop}%
\bibitem [{\citenamefont {Schilcher}\ \emph {et~al.}(2021)\citenamefont
  {Schilcher}, \citenamefont {Robinson}, \citenamefont {Abramovitch},
  \citenamefont {Tan}, \citenamefont {Rappe}, \citenamefont {Reichman},\ and\
  \citenamefont {Egger}}]{schilcher2021significance}%
  \BibitemOpen
  \bibfield  {author} {\bibinfo {author} {\bibfnamefont {M.~J.}\ \bibnamefont
  {Schilcher}}, \bibinfo {author} {\bibfnamefont {P.~J.}\ \bibnamefont
  {Robinson}}, \bibinfo {author} {\bibfnamefont {D.~J.}\ \bibnamefont
  {Abramovitch}}, \bibinfo {author} {\bibfnamefont {L.~Z.}\ \bibnamefont
  {Tan}}, \bibinfo {author} {\bibfnamefont {A.~M.}\ \bibnamefont {Rappe}},
  \bibinfo {author} {\bibfnamefont {D.~R.}\ \bibnamefont {Reichman}}, \ and\
  \bibinfo {author} {\bibfnamefont {D.~A.}\ \bibnamefont {Egger}},\ }\href@noop
  {} {\bibfield  {journal} {\bibinfo  {journal} {ACS Energy Letters}\ }\textbf
  {\bibinfo {volume} {6}},\ \bibinfo {pages} {2162} (\bibinfo {year}
  {2021})}\BibitemShut {NoStop}%
\bibitem [{\citenamefont {Bischak}\ \emph {et~al.}(2017)\citenamefont
  {Bischak}, \citenamefont {Hetherington}, \citenamefont {Wu}, \citenamefont
  {Aloni}, \citenamefont {Ogletree}, \citenamefont {Limmer},\ and\
  \citenamefont {Ginsberg}}]{bischak2017origin}%
  \BibitemOpen
  \bibfield  {author} {\bibinfo {author} {\bibfnamefont {C.~G.}\ \bibnamefont
  {Bischak}}, \bibinfo {author} {\bibfnamefont {C.~L.}\ \bibnamefont
  {Hetherington}}, \bibinfo {author} {\bibfnamefont {H.}~\bibnamefont {Wu}},
  \bibinfo {author} {\bibfnamefont {S.}~\bibnamefont {Aloni}}, \bibinfo
  {author} {\bibfnamefont {D.~F.}\ \bibnamefont {Ogletree}}, \bibinfo {author}
  {\bibfnamefont {D.~T.}\ \bibnamefont {Limmer}}, \ and\ \bibinfo {author}
  {\bibfnamefont {N.~S.}\ \bibnamefont {Ginsberg}},\ }\href@noop {} {\bibfield
  {journal} {\bibinfo  {journal} {Nano letters}\ }\textbf {\bibinfo {volume}
  {17}},\ \bibinfo {pages} {1028} (\bibinfo {year} {2017})}\BibitemShut
  {NoStop}%
\bibitem [{\citenamefont {Mayers}\ \emph {et~al.}(2018)\citenamefont {Mayers},
  \citenamefont {Tan}, \citenamefont {Egger}, \citenamefont {Rappe},\ and\
  \citenamefont {Reichman}}]{mayers2018lattice}%
  \BibitemOpen
  \bibfield  {author} {\bibinfo {author} {\bibfnamefont {M.~Z.}\ \bibnamefont
  {Mayers}}, \bibinfo {author} {\bibfnamefont {L.~Z.}\ \bibnamefont {Tan}},
  \bibinfo {author} {\bibfnamefont {D.~A.}\ \bibnamefont {Egger}}, \bibinfo
  {author} {\bibfnamefont {A.~M.}\ \bibnamefont {Rappe}}, \ and\ \bibinfo
  {author} {\bibfnamefont {D.~R.}\ \bibnamefont {Reichman}},\ }\href@noop {}
  {\bibfield  {journal} {\bibinfo  {journal} {Nano Letters}\ }\textbf {\bibinfo
  {volume} {18}},\ \bibinfo {pages} {8041} (\bibinfo {year}
  {2018})}\BibitemShut {NoStop}%
\bibitem [{\citenamefont {Stranks}\ \emph {et~al.}(2013)\citenamefont
  {Stranks}, \citenamefont {Eperon}, \citenamefont {Grancini}, \citenamefont
  {Menelaou}, \citenamefont {Alcocer}, \citenamefont {Leijtens}, \citenamefont
  {Herz}, \citenamefont {Petrozza},\ and\ \citenamefont
  {Snaith}}]{stranks2013electron}%
  \BibitemOpen
  \bibfield  {author} {\bibinfo {author} {\bibfnamefont {S.~D.}\ \bibnamefont
  {Stranks}}, \bibinfo {author} {\bibfnamefont {G.~E.}\ \bibnamefont {Eperon}},
  \bibinfo {author} {\bibfnamefont {G.}~\bibnamefont {Grancini}}, \bibinfo
  {author} {\bibfnamefont {C.}~\bibnamefont {Menelaou}}, \bibinfo {author}
  {\bibfnamefont {M.~J.}\ \bibnamefont {Alcocer}}, \bibinfo {author}
  {\bibfnamefont {T.}~\bibnamefont {Leijtens}}, \bibinfo {author}
  {\bibfnamefont {L.~M.}\ \bibnamefont {Herz}}, \bibinfo {author}
  {\bibfnamefont {A.}~\bibnamefont {Petrozza}}, \ and\ \bibinfo {author}
  {\bibfnamefont {H.~J.}\ \bibnamefont {Snaith}},\ }\href@noop {} {\bibfield
  {journal} {\bibinfo  {journal} {Science}\ }\textbf {\bibinfo {volume}
  {342}},\ \bibinfo {pages} {341} (\bibinfo {year} {2013})}\BibitemShut
  {NoStop}%
\bibitem [{\citenamefont {Park}\ \emph {et~al.}(2022)\citenamefont {Park},
  \citenamefont {Obliger},\ and\ \citenamefont {Limmer}}]{park2022}%
  \BibitemOpen
  \bibfield  {author} {\bibinfo {author} {\bibfnamefont {Y.}~\bibnamefont
  {Park}}, \bibinfo {author} {\bibfnamefont {A.}~\bibnamefont {Obliger}}, \
  and\ \bibinfo {author} {\bibfnamefont {D.~T.}\ \bibnamefont {Limmer}},\
  }\href {\doibase 10.1021/acs.nanolett.2c00077} {\bibfield  {journal}
  {\bibinfo  {journal} {Nano Letters}\ }\textbf {\bibinfo {volume} {22}},\
  \bibinfo {pages} {2398} (\bibinfo {year} {2022})}\BibitemShut {NoStop}%
\bibitem [{\citenamefont {Martin}\ and\ \citenamefont
  {Frost}(2023)}]{martin2023multiple}%
  \BibitemOpen
  \bibfield  {author} {\bibinfo {author} {\bibfnamefont {B.~A.}\ \bibnamefont
  {Martin}}\ and\ \bibinfo {author} {\bibfnamefont {J.~M.}\ \bibnamefont
  {Frost}},\ }\href@noop {} {\bibfield  {journal} {\bibinfo  {journal}
  {Physical Review B}\ }\textbf {\bibinfo {volume} {107}},\ \bibinfo {pages}
  {115203} (\bibinfo {year} {2023})}\BibitemShut {NoStop}%
\bibitem [{\citenamefont {Zhao}\ \emph {et~al.}(2019)\citenamefont {Zhao},
  \citenamefont {Qin}, \citenamefont {Zhang}, \citenamefont {Wang},
  \citenamefont {Huang}, \citenamefont {Li}, \citenamefont {Dai},\ and\
  \citenamefont {Xiao}}]{10.1021/acs.jpclett.9b00524}%
  \BibitemOpen
  \bibfield  {author} {\bibinfo {author} {\bibfnamefont {W.}~\bibnamefont
  {Zhao}}, \bibinfo {author} {\bibfnamefont {Z.}~\bibnamefont {Qin}}, \bibinfo
  {author} {\bibfnamefont {C.}~\bibnamefont {Zhang}}, \bibinfo {author}
  {\bibfnamefont {G.}~\bibnamefont {Wang}}, \bibinfo {author} {\bibfnamefont
  {X.}~\bibnamefont {Huang}}, \bibinfo {author} {\bibfnamefont
  {B.}~\bibnamefont {Li}}, \bibinfo {author} {\bibfnamefont {X.}~\bibnamefont
  {Dai}}, \ and\ \bibinfo {author} {\bibfnamefont {M.}~\bibnamefont {Xiao}},\
  }\href {\doibase 10.1021/acs.jpclett.9b00524} {\bibfield  {journal} {\bibinfo
   {journal} {The Journal of Physical Chemistry Letters}\ }\textbf {\bibinfo
  {volume} {10}},\ \bibinfo {pages} {1251} (\bibinfo {year}
  {2019})}\BibitemShut {NoStop}%
\bibitem [{\citenamefont {Dana}\ \emph {et~al.}(2021)\citenamefont {Dana},
  \citenamefont {Binyamin}, \citenamefont {Etgar},\ and\ \citenamefont
  {Ruhman}}]{dana2021unusually}%
  \BibitemOpen
  \bibfield  {author} {\bibinfo {author} {\bibfnamefont {J.}~\bibnamefont
  {Dana}}, \bibinfo {author} {\bibfnamefont {T.}~\bibnamefont {Binyamin}},
  \bibinfo {author} {\bibfnamefont {L.}~\bibnamefont {Etgar}}, \ and\ \bibinfo
  {author} {\bibfnamefont {S.}~\bibnamefont {Ruhman}},\ }\href@noop {}
  {\bibfield  {journal} {\bibinfo  {journal} {ACS Nano}\ }\textbf {\bibinfo
  {volume} {15}},\ \bibinfo {pages} {9039} (\bibinfo {year}
  {2021})}\BibitemShut {NoStop}%
\bibitem [{\citenamefont {Protesescu}\ \emph
  {et~al.}(2015{\natexlab{a}})\citenamefont {Protesescu}, \citenamefont
  {Yakunin}, \citenamefont {Bodnarchuk}, \citenamefont {Krieg}, \citenamefont
  {Caputo}, \citenamefont {Hendon}, \citenamefont {Yang}, \citenamefont
  {Walsh},\ and\ \citenamefont {Kovalenko}}]{10.1021/nl5048779}%
  \BibitemOpen
  \bibfield  {author} {\bibinfo {author} {\bibfnamefont {L.}~\bibnamefont
  {Protesescu}}, \bibinfo {author} {\bibfnamefont {S.}~\bibnamefont {Yakunin}},
  \bibinfo {author} {\bibfnamefont {M.~I.}\ \bibnamefont {Bodnarchuk}},
  \bibinfo {author} {\bibfnamefont {F.}~\bibnamefont {Krieg}}, \bibinfo
  {author} {\bibfnamefont {R.}~\bibnamefont {Caputo}}, \bibinfo {author}
  {\bibfnamefont {C.~H.}\ \bibnamefont {Hendon}}, \bibinfo {author}
  {\bibfnamefont {R.~X.}\ \bibnamefont {Yang}}, \bibinfo {author}
  {\bibfnamefont {A.}~\bibnamefont {Walsh}}, \ and\ \bibinfo {author}
  {\bibfnamefont {M.~V.}\ \bibnamefont {Kovalenko}},\ }\href {\doibase
  10.1021/nl5048779} {\bibfield  {journal} {\bibinfo  {journal} {Nano Letters}\
  }\textbf {\bibinfo {volume} {15}},\ \bibinfo {pages} {3692} (\bibinfo {year}
  {2015}{\natexlab{a}})}\BibitemShut {NoStop}%
\bibitem [{\citenamefont {Parveen}\ \emph {et~al.}(2019)\citenamefont
  {Parveen}, \citenamefont {Paul}, \citenamefont {Das},\ and\ \citenamefont
  {Giri}}]{10.1016/j.jcis.2018.12.105}%
  \BibitemOpen
  \bibfield  {author} {\bibinfo {author} {\bibfnamefont {S.}~\bibnamefont
  {Parveen}}, \bibinfo {author} {\bibfnamefont {K.~K.}\ \bibnamefont {Paul}},
  \bibinfo {author} {\bibfnamefont {R.}~\bibnamefont {Das}}, \ and\ \bibinfo
  {author} {\bibfnamefont {P.}~\bibnamefont {Giri}},\ }\href {\doibase
  10.1016/j.jcis.2018.12.105} {\bibfield  {journal} {\bibinfo  {journal}
  {Journal of Colloid and Interface Science}\ }\textbf {\bibinfo {volume}
  {539}},\ \bibinfo {pages} {619} (\bibinfo {year} {2019})}\BibitemShut
  {NoStop}%
\bibitem [{\citenamefont {Yang}\ and\ \citenamefont
  {Tan}(2020)}]{10.1063/1.5128016}%
  \BibitemOpen
  \bibfield  {author} {\bibinfo {author} {\bibfnamefont {R.~X.}\ \bibnamefont
  {Yang}}\ and\ \bibinfo {author} {\bibfnamefont {L.~Z.}\ \bibnamefont {Tan}},\
  }\href {\doibase 10.1063/1.5128016} {\bibfield  {journal} {\bibinfo
  {journal} {The Journal of Chemical Physics}\ }\textbf {\bibinfo {volume}
  {152}},\ \bibinfo {pages} {034702} (\bibinfo {year} {2020})}\BibitemShut
  {NoStop}%
\bibitem [{\citenamefont {Li}\ \emph {et~al.}(2017)\citenamefont {Li},
  \citenamefont {Luo}, \citenamefont {Huang}, \citenamefont {Ma}, \citenamefont
  {Ye}, \citenamefont {Zeng},\ and\ \citenamefont
  {He}}]{10.1021/acs.jpclett.7b00017}%
  \BibitemOpen
  \bibfield  {author} {\bibinfo {author} {\bibfnamefont {J.}~\bibnamefont
  {Li}}, \bibinfo {author} {\bibfnamefont {L.}~\bibnamefont {Luo}}, \bibinfo
  {author} {\bibfnamefont {H.}~\bibnamefont {Huang}}, \bibinfo {author}
  {\bibfnamefont {C.}~\bibnamefont {Ma}}, \bibinfo {author} {\bibfnamefont
  {Z.}~\bibnamefont {Ye}}, \bibinfo {author} {\bibfnamefont {J.}~\bibnamefont
  {Zeng}}, \ and\ \bibinfo {author} {\bibfnamefont {H.}~\bibnamefont {He}},\
  }\href {\doibase 10.1021/acs.jpclett.7b00017} {\bibfield  {journal} {\bibinfo
   {journal} {The Journal of Physical Chemistry Letters}\ }\textbf {\bibinfo
  {volume} {8}},\ \bibinfo {pages} {1161} (\bibinfo {year} {2017})}\BibitemShut
  {NoStop}%
\bibitem [{\citenamefont {Zheng}\ \emph {et~al.}(2015)\citenamefont {Zheng},
  \citenamefont {Zhu}, \citenamefont {Abdellah}, \citenamefont {Messing},
  \citenamefont {Zhang}, \citenamefont {Generalov}, \citenamefont {Niu},
  \citenamefont {Ribaud}, \citenamefont {Canton},\ and\ \citenamefont
  {Pullerits}}]{10.1021/acs.jpclett.5b01252}%
  \BibitemOpen
  \bibfield  {author} {\bibinfo {author} {\bibfnamefont {K.}~\bibnamefont
  {Zheng}}, \bibinfo {author} {\bibfnamefont {Q.}~\bibnamefont {Zhu}}, \bibinfo
  {author} {\bibfnamefont {M.}~\bibnamefont {Abdellah}}, \bibinfo {author}
  {\bibfnamefont {M.~E.}\ \bibnamefont {Messing}}, \bibinfo {author}
  {\bibfnamefont {W.}~\bibnamefont {Zhang}}, \bibinfo {author} {\bibfnamefont
  {A.}~\bibnamefont {Generalov}}, \bibinfo {author} {\bibfnamefont
  {Y.}~\bibnamefont {Niu}}, \bibinfo {author} {\bibfnamefont {L.}~\bibnamefont
  {Ribaud}}, \bibinfo {author} {\bibfnamefont {S.~E.}\ \bibnamefont {Canton}},
  \ and\ \bibinfo {author} {\bibfnamefont {T.}~\bibnamefont {Pullerits}},\
  }\href {\doibase 10.1021/acs.jpclett.5b01252} {\bibfield  {journal} {\bibinfo
   {journal} {The Journal of Physical Chemistry Letters}\ }\textbf {\bibinfo
  {volume} {6}},\ \bibinfo {pages} {2969} (\bibinfo {year} {2015})}\BibitemShut
  {NoStop}%
\bibitem [{\citenamefont {Qaid}\ \emph {et~al.}(2021)\citenamefont {Qaid},
  \citenamefont {Ghaithan}, \citenamefont {Al-Asbahi},\ and\ \citenamefont
  {Aldwayyan}}]{10.1021/acsomega.0c05414}%
  \BibitemOpen
  \bibfield  {author} {\bibinfo {author} {\bibfnamefont {S.~M.~H.}\
  \bibnamefont {Qaid}}, \bibinfo {author} {\bibfnamefont {H.~M.}\ \bibnamefont
  {Ghaithan}}, \bibinfo {author} {\bibfnamefont {B.~A.}\ \bibnamefont
  {Al-Asbahi}}, \ and\ \bibinfo {author} {\bibfnamefont {A.~S.}\ \bibnamefont
  {Aldwayyan}},\ }\href {\doibase 10.1021/acsomega.0c05414} {\bibfield
  {journal} {\bibinfo  {journal} {ACS Omega}\ }\textbf {\bibinfo {volume}
  {6}},\ \bibinfo {pages} {5297} (\bibinfo {year} {2021})}\BibitemShut
  {NoStop}%
\bibitem [{\citenamefont {Ashner}\ \emph {et~al.}(2019)\citenamefont {Ashner},
  \citenamefont {Shulenberger}, \citenamefont {Krieg}, \citenamefont {Powers},
  \citenamefont {Kovalenko}, \citenamefont {Bawendi},\ and\ \citenamefont
  {Tisdale}}]{10.1021/acsenergylett.9b02041}%
  \BibitemOpen
  \bibfield  {author} {\bibinfo {author} {\bibfnamefont {M.~N.}\ \bibnamefont
  {Ashner}}, \bibinfo {author} {\bibfnamefont {K.~E.}\ \bibnamefont
  {Shulenberger}}, \bibinfo {author} {\bibfnamefont {F.}~\bibnamefont {Krieg}},
  \bibinfo {author} {\bibfnamefont {E.~R.}\ \bibnamefont {Powers}}, \bibinfo
  {author} {\bibfnamefont {M.~V.}\ \bibnamefont {Kovalenko}}, \bibinfo {author}
  {\bibfnamefont {M.~G.}\ \bibnamefont {Bawendi}}, \ and\ \bibinfo {author}
  {\bibfnamefont {W.~A.}\ \bibnamefont {Tisdale}},\ }\href {\doibase
  10.1021/acsenergylett.9b02041} {\bibfield  {journal} {\bibinfo  {journal}
  {ACS Energy Letters}\ }\textbf {\bibinfo {volume} {4}},\ \bibinfo {pages}
  {2639} (\bibinfo {year} {2019})}\BibitemShut {NoStop}%
\bibitem [{\citenamefont {Amara}\ \emph {et~al.}(2023)\citenamefont {Amara},
  \citenamefont {Said}, \citenamefont {Huo}, \citenamefont {Pierret},
  \citenamefont {Voisin}, \citenamefont {Gao}, \citenamefont {Xiong},\ and\
  \citenamefont {Diederichs}}]{10.1021/acs.nanolett.3c00793}%
  \BibitemOpen
  \bibfield  {author} {\bibinfo {author} {\bibfnamefont {M.-R.}\ \bibnamefont
  {Amara}}, \bibinfo {author} {\bibfnamefont {Z.}~\bibnamefont {Said}},
  \bibinfo {author} {\bibfnamefont {C.}~\bibnamefont {Huo}}, \bibinfo {author}
  {\bibfnamefont {A.}~\bibnamefont {Pierret}}, \bibinfo {author} {\bibfnamefont
  {C.}~\bibnamefont {Voisin}}, \bibinfo {author} {\bibfnamefont
  {W.}~\bibnamefont {Gao}}, \bibinfo {author} {\bibfnamefont {Q.}~\bibnamefont
  {Xiong}}, \ and\ \bibinfo {author} {\bibfnamefont {C.}~\bibnamefont
  {Diederichs}},\ }\href {\doibase 10.1021/acs.nanolett.3c00793} {\bibfield
  {journal} {\bibinfo  {journal} {Nano Letters}\ }\textbf {\bibinfo {volume}
  {23}} (\bibinfo {year} {2023}),\ 10.1021/acs.nanolett.3c00793},\ \Eprint
  {http://arxiv.org/abs/2301.09571} {2301.09571} \BibitemShut {NoStop}%
\bibitem [{\citenamefont {Makarov}\ \emph {et~al.}(2016)\citenamefont
  {Makarov}, \citenamefont {Guo}, \citenamefont {Isaienko}, \citenamefont
  {Liu}, \citenamefont {Robel},\ and\ \citenamefont
  {Klimov}}]{10.1021/acs.nanolett.5b05077}%
  \BibitemOpen
  \bibfield  {author} {\bibinfo {author} {\bibfnamefont {N.~S.}\ \bibnamefont
  {Makarov}}, \bibinfo {author} {\bibfnamefont {S.}~\bibnamefont {Guo}},
  \bibinfo {author} {\bibfnamefont {O.}~\bibnamefont {Isaienko}}, \bibinfo
  {author} {\bibfnamefont {W.}~\bibnamefont {Liu}}, \bibinfo {author}
  {\bibfnamefont {I.}~\bibnamefont {Robel}}, \ and\ \bibinfo {author}
  {\bibfnamefont {V.~I.}\ \bibnamefont {Klimov}},\ }\href {\doibase
  10.1021/acs.nanolett.5b05077} {\bibfield  {journal} {\bibinfo  {journal}
  {Nano Letters}\ }\textbf {\bibinfo {volume} {16}},\ \bibinfo {pages} {2349}
  (\bibinfo {year} {2016})}\BibitemShut {NoStop}%
\bibitem [{\citenamefont {Cho}\ \emph {et~al.}(2021{\natexlab{a}})\citenamefont
  {Cho}, \citenamefont {Yamada}, \citenamefont {Tahara}, \citenamefont
  {Tadano}, \citenamefont {Suzuura}, \citenamefont {Saruyama}, \citenamefont
  {Sato}, \citenamefont {Teranishi},\ and\ \citenamefont
  {Kanemitsu}}]{10.1021/acs.nanolett.1c02122}%
  \BibitemOpen
  \bibfield  {author} {\bibinfo {author} {\bibfnamefont {K.}~\bibnamefont
  {Cho}}, \bibinfo {author} {\bibfnamefont {T.}~\bibnamefont {Yamada}},
  \bibinfo {author} {\bibfnamefont {H.}~\bibnamefont {Tahara}}, \bibinfo
  {author} {\bibfnamefont {T.}~\bibnamefont {Tadano}}, \bibinfo {author}
  {\bibfnamefont {H.}~\bibnamefont {Suzuura}}, \bibinfo {author} {\bibfnamefont
  {M.}~\bibnamefont {Saruyama}}, \bibinfo {author} {\bibfnamefont
  {R.}~\bibnamefont {Sato}}, \bibinfo {author} {\bibfnamefont {T.}~\bibnamefont
  {Teranishi}}, \ and\ \bibinfo {author} {\bibfnamefont {Y.}~\bibnamefont
  {Kanemitsu}},\ }\href {\doibase 10.1021/acs.nanolett.1c02122} {\bibfield
  {journal} {\bibinfo  {journal} {Nano Letters}\ }\textbf {\bibinfo {volume}
  {21}},\ \bibinfo {pages} {7206} (\bibinfo {year}
  {2021}{\natexlab{a}})}\BibitemShut {NoStop}%
\bibitem [{\citenamefont {Yumoto}\ \emph {et~al.}(2018)\citenamefont {Yumoto},
  \citenamefont {Tahara}, \citenamefont {Kawawaki}, \citenamefont {Saruyama},
  \citenamefont {Sato}, \citenamefont {Teranishi},\ and\ \citenamefont
  {Kanemitsu}}]{10.1021/acs.jpclett.8b01029}%
  \BibitemOpen
  \bibfield  {author} {\bibinfo {author} {\bibfnamefont {G.}~\bibnamefont
  {Yumoto}}, \bibinfo {author} {\bibfnamefont {H.}~\bibnamefont {Tahara}},
  \bibinfo {author} {\bibfnamefont {T.}~\bibnamefont {Kawawaki}}, \bibinfo
  {author} {\bibfnamefont {M.}~\bibnamefont {Saruyama}}, \bibinfo {author}
  {\bibfnamefont {R.}~\bibnamefont {Sato}}, \bibinfo {author} {\bibfnamefont
  {T.}~\bibnamefont {Teranishi}}, \ and\ \bibinfo {author} {\bibfnamefont
  {Y.}~\bibnamefont {Kanemitsu}},\ }\href {\doibase
  10.1021/acs.jpclett.8b01029} {\bibfield  {journal} {\bibinfo  {journal} {The
  Journal of Physical Chemistry Letters}\ }\textbf {\bibinfo {volume} {9}},\
  \bibinfo {pages} {2222} (\bibinfo {year} {2018})}\BibitemShut {NoStop}%
\bibitem [{\citenamefont {Aneesh}\ \emph {et~al.}(2017)\citenamefont {Aneesh},
  \citenamefont {Swarnkar}, \citenamefont {Ravi}, \citenamefont {Sharma},
  \citenamefont {Nag},\ and\ \citenamefont
  {Adarsh}}]{10.1021/acs.jpcc.7b00762}%
  \BibitemOpen
  \bibfield  {author} {\bibinfo {author} {\bibfnamefont {J.}~\bibnamefont
  {Aneesh}}, \bibinfo {author} {\bibfnamefont {A.}~\bibnamefont {Swarnkar}},
  \bibinfo {author} {\bibfnamefont {V.~K.}\ \bibnamefont {Ravi}}, \bibinfo
  {author} {\bibfnamefont {R.}~\bibnamefont {Sharma}}, \bibinfo {author}
  {\bibfnamefont {A.}~\bibnamefont {Nag}}, \ and\ \bibinfo {author}
  {\bibfnamefont {K.~V.}\ \bibnamefont {Adarsh}},\ }\href@noop {} {\bibfield
  {journal} {\bibinfo  {journal} {The Journal of Physical Chemistry C}\
  }\textbf {\bibinfo {volume} {121}},\ \bibinfo {pages} {4734} (\bibinfo {year}
  {2017})}\BibitemShut {NoStop}%
\bibitem [{\citenamefont {Castaneda}\ \emph {et~al.}(2016)\citenamefont
  {Castaneda}, \citenamefont {Nagamine}, \citenamefont {Yassitepe},
  \citenamefont {Bonato}, \citenamefont {Voznyy}, \citenamefont {Hoogland},
  \citenamefont {Nogueira}, \citenamefont {Sargent}, \citenamefont {Cruz},\
  and\ \citenamefont {Padilha}}]{10.1021/acsnano.6b03908}%
  \BibitemOpen
  \bibfield  {author} {\bibinfo {author} {\bibfnamefont {J.~A.}\ \bibnamefont
  {Castaneda}}, \bibinfo {author} {\bibfnamefont {G.}~\bibnamefont {Nagamine}},
  \bibinfo {author} {\bibfnamefont {E.}~\bibnamefont {Yassitepe}}, \bibinfo
  {author} {\bibfnamefont {L.~G.}\ \bibnamefont {Bonato}}, \bibinfo {author}
  {\bibfnamefont {O.}~\bibnamefont {Voznyy}}, \bibinfo {author} {\bibfnamefont
  {S.}~\bibnamefont {Hoogland}}, \bibinfo {author} {\bibfnamefont {A.~F.}\
  \bibnamefont {Nogueira}}, \bibinfo {author} {\bibfnamefont {E.~H.}\
  \bibnamefont {Sargent}}, \bibinfo {author} {\bibfnamefont {C.~H.~B.}\
  \bibnamefont {Cruz}}, \ and\ \bibinfo {author} {\bibfnamefont {L.~A.}\
  \bibnamefont {Padilha}},\ }\href@noop {} {\bibfield  {journal} {\bibinfo
  {journal} {ACS Nano}\ }\textbf {\bibinfo {volume} {10}},\ \bibinfo {pages}
  {8603} (\bibinfo {year} {2016})}\BibitemShut {NoStop}%
\bibitem [{\citenamefont {Lubin}\ \emph {et~al.}(2021)\citenamefont {Lubin},
  \citenamefont {Tenne}, \citenamefont {Ulku}, \citenamefont {Antolovic},
  \citenamefont {Burri}, \citenamefont {Karg}, \citenamefont {Yallapragada},
  \citenamefont {Bruschini}, \citenamefont {Charbon},\ and\ \citenamefont
  {Oron}}]{10.1021/acs.nanolett.1c01291}%
  \BibitemOpen
  \bibfield  {author} {\bibinfo {author} {\bibfnamefont {G.}~\bibnamefont
  {Lubin}}, \bibinfo {author} {\bibfnamefont {R.}~\bibnamefont {Tenne}},
  \bibinfo {author} {\bibfnamefont {A.~C.}\ \bibnamefont {Ulku}}, \bibinfo
  {author} {\bibfnamefont {I.~M.}\ \bibnamefont {Antolovic}}, \bibinfo {author}
  {\bibfnamefont {S.}~\bibnamefont {Burri}}, \bibinfo {author} {\bibfnamefont
  {S.}~\bibnamefont {Karg}}, \bibinfo {author} {\bibfnamefont {V.~J.}\
  \bibnamefont {Yallapragada}}, \bibinfo {author} {\bibfnamefont
  {C.}~\bibnamefont {Bruschini}}, \bibinfo {author} {\bibfnamefont
  {E.}~\bibnamefont {Charbon}}, \ and\ \bibinfo {author} {\bibfnamefont
  {D.}~\bibnamefont {Oron}},\ }\href {\doibase 10.1021/acs.nanolett.1c01291}
  {\bibfield  {journal} {\bibinfo  {journal} {Nano Letters}\ }\textbf {\bibinfo
  {volume} {21}},\ \bibinfo {pages} {6756} (\bibinfo {year} {2021})},\ \Eprint
  {http://arxiv.org/abs/2108.00345} {2108.00345} \BibitemShut {NoStop}%
\bibitem [{\citenamefont {Antolinez}\ \emph {et~al.}(2019)\citenamefont
  {Antolinez}, \citenamefont {Rabouw}, \citenamefont {Rossinelli},
  \citenamefont {Cui},\ and\ \citenamefont
  {Norris}}]{10.1021/acs.nanolett.9b02856}%
  \BibitemOpen
  \bibfield  {author} {\bibinfo {author} {\bibfnamefont {F.~V.}\ \bibnamefont
  {Antolinez}}, \bibinfo {author} {\bibfnamefont {F.~T.}\ \bibnamefont
  {Rabouw}}, \bibinfo {author} {\bibfnamefont {A.~A.}\ \bibnamefont
  {Rossinelli}}, \bibinfo {author} {\bibfnamefont {J.}~\bibnamefont {Cui}}, \
  and\ \bibinfo {author} {\bibfnamefont {D.~J.}\ \bibnamefont {Norris}},\
  }\href {\doibase 10.1021/acs.nanolett.9b02856} {\bibfield  {journal}
  {\bibinfo  {journal} {Nano Letters}\ }\textbf {\bibinfo {volume} {19}},\
  \bibinfo {pages} {8495} (\bibinfo {year} {2019})}\BibitemShut {NoStop}%
\bibitem [{\citenamefont {Beyler}\ \emph {et~al.}(2013)\citenamefont {Beyler},
  \citenamefont {Marshall}, \citenamefont {Cui}, \citenamefont {Brokmann},\
  and\ \citenamefont {Bawendi}}]{10.1103/physrevlett.111.177401}%
  \BibitemOpen
  \bibfield  {author} {\bibinfo {author} {\bibfnamefont {A.~P.}\ \bibnamefont
  {Beyler}}, \bibinfo {author} {\bibfnamefont {L.~F.}\ \bibnamefont
  {Marshall}}, \bibinfo {author} {\bibfnamefont {J.}~\bibnamefont {Cui}},
  \bibinfo {author} {\bibfnamefont {X.}~\bibnamefont {Brokmann}}, \ and\
  \bibinfo {author} {\bibfnamefont {M.~G.}\ \bibnamefont {Bawendi}},\ }\href
  {\doibase 10.1103/physrevlett.111.177401} {\bibfield  {journal} {\bibinfo
  {journal} {Physical Review Letters}\ }\textbf {\bibinfo {volume} {111}},\
  \bibinfo {pages} {177401} (\bibinfo {year} {2013})}\BibitemShut {NoStop}%
\bibitem [{\citenamefont {Pollmann}\ and\ \citenamefont
  {B{\"u}ttner}(1977)}]{pollmann1977effective}%
  \BibitemOpen
  \bibfield  {author} {\bibinfo {author} {\bibfnamefont {J.}~\bibnamefont
  {Pollmann}}\ and\ \bibinfo {author} {\bibfnamefont {H.}~\bibnamefont
  {B{\"u}ttner}},\ }\href@noop {} {\bibfield  {journal} {\bibinfo  {journal}
  {Physical Review B}\ }\textbf {\bibinfo {volume} {16}},\ \bibinfo {pages}
  {4480} (\bibinfo {year} {1977})}\BibitemShut {NoStop}%
\bibitem [{\citenamefont {hook}(1956)}]{haken1956quantum}%
  \BibitemOpen
  \bibfield  {author} {\bibinfo {author} {\bibfnamefont {H.}~\bibnamefont
  {hook}},\ }\href@noop {} {\bibfield  {journal} {\bibinfo  {journal} {journal
  for physics}\ }\textbf {\bibinfo {volume} {146}},\ \bibinfo {pages} {527}
  (\bibinfo {year} {1956})}\BibitemShut {NoStop}%
\bibitem [{\citenamefont {Park}\ and\ \citenamefont
  {Limmer}(2022)}]{parkjcp2022}%
  \BibitemOpen
  \bibfield  {author} {\bibinfo {author} {\bibfnamefont {Y.}~\bibnamefont
  {Park}}\ and\ \bibinfo {author} {\bibfnamefont {D.~T.}\ \bibnamefont
  {Limmer}},\ }\href {\doibase 10.1063/5.0100738} {\bibfield  {journal}
  {\bibinfo  {journal} {The Journal of Chemical Physics}\ }\textbf {\bibinfo
  {volume} {157}},\ \bibinfo {pages} {104116} (\bibinfo {year} {2022})},\
  \Eprint {http://arxiv.org/abs/2205.11780} {2205.11780} \BibitemShut {NoStop}%
\bibitem [{\citenamefont {Filip}\ \emph {et~al.}(2021)\citenamefont {Filip},
  \citenamefont {Haber},\ and\ \citenamefont {Neaton}}]{neatonprl}%
  \BibitemOpen
  \bibfield  {author} {\bibinfo {author} {\bibfnamefont {M.~R.}\ \bibnamefont
  {Filip}}, \bibinfo {author} {\bibfnamefont {J.~B.}\ \bibnamefont {Haber}}, \
  and\ \bibinfo {author} {\bibfnamefont {J.~B.}\ \bibnamefont {Neaton}},\
  }\href {\doibase 10.1103/physrevlett.127.067401} {\bibfield  {journal}
  {\bibinfo  {journal} {Physical Review Letters}\ }\textbf {\bibinfo {volume}
  {127}},\ \bibinfo {pages} {067401} (\bibinfo {year} {2021})},\ \Eprint
  {http://arxiv.org/abs/2106.08697} {2106.08697} \BibitemShut {NoStop}%
\bibitem [{\citenamefont {Schlipf}\ \emph {et~al.}(2018)\citenamefont
  {Schlipf}, \citenamefont {Poncé},\ and\ \citenamefont
  {Giustino}}]{multiphonon}%
  \BibitemOpen
  \bibfield  {author} {\bibinfo {author} {\bibfnamefont {M.}~\bibnamefont
  {Schlipf}}, \bibinfo {author} {\bibfnamefont {S.}~\bibnamefont {Poncé}}, \
  and\ \bibinfo {author} {\bibfnamefont {F.}~\bibnamefont {Giustino}},\ }\href
  {\doibase 10.1103/physrevlett.121.086402} {\bibfield  {journal} {\bibinfo
  {journal} {Physical Review Letters}\ }\textbf {\bibinfo {volume} {121}},\
  \bibinfo {pages} {086402} (\bibinfo {year} {2018})},\ \Eprint
  {http://arxiv.org/abs/1808.08130} {1808.08130} \BibitemShut {NoStop}%
\bibitem [{\citenamefont {Sio}\ \emph {et~al.}(2019)\citenamefont {Sio},
  \citenamefont {Verdi}, \citenamefont {Poncé},\ and\ \citenamefont
  {Giustino}}]{giustinoPRB}%
  \BibitemOpen
  \bibfield  {author} {\bibinfo {author} {\bibfnamefont {W.~H.}\ \bibnamefont
  {Sio}}, \bibinfo {author} {\bibfnamefont {C.}~\bibnamefont {Verdi}}, \bibinfo
  {author} {\bibfnamefont {S.}~\bibnamefont {Poncé}}, \ and\ \bibinfo {author}
  {\bibfnamefont {F.}~\bibnamefont {Giustino}},\ }\href {\doibase
  10.1103/physrevb.99.235139} {\bibfield  {journal} {\bibinfo  {journal}
  {Physical Review B}\ }\textbf {\bibinfo {volume} {99}},\ \bibinfo {pages}
  {235139} (\bibinfo {year} {2019})},\ \Eprint
  {http://arxiv.org/abs/1906.08408} {1906.08408} \BibitemShut {NoStop}%
\bibitem [{\citenamefont {Sio}\ and\ \citenamefont
  {Giustino}(2023)}]{giustinopolaron}%
  \BibitemOpen
  \bibfield  {author} {\bibinfo {author} {\bibfnamefont {W.~H.}\ \bibnamefont
  {Sio}}\ and\ \bibinfo {author} {\bibfnamefont {F.}~\bibnamefont {Giustino}},\
  }\href@noop {} {\bibfield  {journal} {\bibinfo  {journal} {Nature Physics}\
  }\textbf {\bibinfo {volume} {19}},\ \bibinfo {pages} {629} (\bibinfo {year}
  {2023})}\BibitemShut {NoStop}%
\bibitem [{\citenamefont {Rohlfing}\ and\ \citenamefont
  {Louie}(2000)}]{rohlfing2000electron}%
  \BibitemOpen
  \bibfield  {author} {\bibinfo {author} {\bibfnamefont {M.}~\bibnamefont
  {Rohlfing}}\ and\ \bibinfo {author} {\bibfnamefont {S.~G.}\ \bibnamefont
  {Louie}},\ }\href@noop {} {\bibfield  {journal} {\bibinfo  {journal}
  {Physical Review B}\ }\textbf {\bibinfo {volume} {62}},\ \bibinfo {pages}
  {4927} (\bibinfo {year} {2000})}\BibitemShut {NoStop}%
\bibitem [{\citenamefont {Albrecht}\ \emph {et~al.}(1998)\citenamefont
  {Albrecht}, \citenamefont {Reining}, \citenamefont {Del~Sole},\ and\
  \citenamefont {Onida}}]{albrecht1998excitonic}%
  \BibitemOpen
  \bibfield  {author} {\bibinfo {author} {\bibfnamefont {S.}~\bibnamefont
  {Albrecht}}, \bibinfo {author} {\bibfnamefont {L.}~\bibnamefont {Reining}},
  \bibinfo {author} {\bibfnamefont {R.}~\bibnamefont {Del~Sole}}, \ and\
  \bibinfo {author} {\bibfnamefont {G.}~\bibnamefont {Onida}},\ }\href@noop {}
  {\bibfield  {journal} {\bibinfo  {journal} {physica status solidi (a)}\
  }\textbf {\bibinfo {volume} {170}},\ \bibinfo {pages} {189} (\bibinfo {year}
  {1998})}\BibitemShut {NoStop}%
\bibitem [{\citenamefont {Hedin}(1965)}]{hedin1965new}%
  \BibitemOpen
  \bibfield  {author} {\bibinfo {author} {\bibfnamefont {L.}~\bibnamefont
  {Hedin}},\ }\href@noop {} {\bibfield  {journal} {\bibinfo  {journal}
  {Physical Review}\ }\textbf {\bibinfo {volume} {139}},\ \bibinfo {pages}
  {A796} (\bibinfo {year} {1965})}\BibitemShut {NoStop}%
\bibitem [{\citenamefont {Cho}\ and\ \citenamefont
  {Berkelbach}(2019)}]{cho2019optical}%
  \BibitemOpen
  \bibfield  {author} {\bibinfo {author} {\bibfnamefont {Y.}~\bibnamefont
  {Cho}}\ and\ \bibinfo {author} {\bibfnamefont {T.~C.}\ \bibnamefont
  {Berkelbach}},\ }\href@noop {} {\bibfield  {journal} {\bibinfo  {journal}
  {The journal of physical chemistry letters}\ }\textbf {\bibinfo {volume}
  {10}},\ \bibinfo {pages} {6189} (\bibinfo {year} {2019})}\BibitemShut
  {NoStop}%
\bibitem [{\citenamefont {Biffi}\ \emph {et~al.}(2023)\citenamefont {Biffi},
  \citenamefont {Cho}, \citenamefont {Krahne},\ and\ \citenamefont
  {Berkelbach}}]{biffi2023excitons}%
  \BibitemOpen
  \bibfield  {author} {\bibinfo {author} {\bibfnamefont {G.}~\bibnamefont
  {Biffi}}, \bibinfo {author} {\bibfnamefont {Y.}~\bibnamefont {Cho}}, \bibinfo
  {author} {\bibfnamefont {R.}~\bibnamefont {Krahne}}, \ and\ \bibinfo {author}
  {\bibfnamefont {T.~C.}\ \bibnamefont {Berkelbach}},\ }\href@noop {}
  {\bibfield  {journal} {\bibinfo  {journal} {The Journal of Physical Chemistry
  C}\ }\textbf {\bibinfo {volume} {127}},\ \bibinfo {pages} {1891} (\bibinfo
  {year} {2023})}\BibitemShut {NoStop}%
\bibitem [{\citenamefont {Cho}\ \emph {et~al.}(2021{\natexlab{b}})\citenamefont
  {Cho}, \citenamefont {Greene},\ and\ \citenamefont
  {Berkelbach}}]{cho2021simulations}%
  \BibitemOpen
  \bibfield  {author} {\bibinfo {author} {\bibfnamefont {Y.}~\bibnamefont
  {Cho}}, \bibinfo {author} {\bibfnamefont {S.~M.}\ \bibnamefont {Greene}}, \
  and\ \bibinfo {author} {\bibfnamefont {T.~C.}\ \bibnamefont {Berkelbach}},\
  }\href@noop {} {\bibfield  {journal} {\bibinfo  {journal} {Physical Review
  Letters}\ }\textbf {\bibinfo {volume} {126}},\ \bibinfo {pages} {216402}
  (\bibinfo {year} {2021}{\natexlab{b}})}\BibitemShut {NoStop}%
\bibitem [{\citenamefont {Weinberg}\ \emph {et~al.}(2023)\citenamefont
  {Weinberg}, \citenamefont {Park}, \citenamefont {Limmer},\ and\ \citenamefont
  {Rabani}}]{danielarxiv}%
  \BibitemOpen
  \bibfield  {author} {\bibinfo {author} {\bibfnamefont {D.}~\bibnamefont
  {Weinberg}}, \bibinfo {author} {\bibfnamefont {Y.}~\bibnamefont {Park}},
  \bibinfo {author} {\bibfnamefont {D.~T.}\ \bibnamefont {Limmer}}, \ and\
  \bibinfo {author} {\bibfnamefont {E.}~\bibnamefont {Rabani}},\ }\href
  {\doibase 10.1021/acs.nanolett.3c00861} {\bibfield  {journal} {\bibinfo
  {journal} {Nano Letters}\ }\textbf {\bibinfo {volume} {23}},\ \bibinfo
  {pages} {4997} (\bibinfo {year} {2023})},\ \bibinfo {note} {pMID:
  37229762}\BibitemShut {NoStop}%
\bibitem [{\citenamefont {Parrinello}\ and\ \citenamefont
  {Rahman}(1984)}]{parrinello1984study}%
  \BibitemOpen
  \bibfield  {author} {\bibinfo {author} {\bibfnamefont {M.}~\bibnamefont
  {Parrinello}}\ and\ \bibinfo {author} {\bibfnamefont {A.}~\bibnamefont
  {Rahman}},\ }\href@noop {} {\bibfield  {journal} {\bibinfo  {journal} {The
  Journal of chemical physics}\ }\textbf {\bibinfo {volume} {80}},\ \bibinfo
  {pages} {860} (\bibinfo {year} {1984})}\BibitemShut {NoStop}%
\bibitem [{\citenamefont {Bischak}\ \emph {et~al.}(2020)\citenamefont
  {Bischak}, \citenamefont {Lai}, \citenamefont {Fan}, \citenamefont {Lu},
  \citenamefont {David}, \citenamefont {Dong}, \citenamefont {Chen},
  \citenamefont {Etman}, \citenamefont {Lei}, \citenamefont {Sun},
  \citenamefont {Grünwald}, \citenamefont {Limmer}, \citenamefont {Yang},\
  and\ \citenamefont {Ginsberg}}]{10.1016/j.matt.2020.07.015}%
  \BibitemOpen
  \bibfield  {author} {\bibinfo {author} {\bibfnamefont {C.~G.}\ \bibnamefont
  {Bischak}}, \bibinfo {author} {\bibfnamefont {M.}~\bibnamefont {Lai}},
  \bibinfo {author} {\bibfnamefont {Z.}~\bibnamefont {Fan}}, \bibinfo {author}
  {\bibfnamefont {D.}~\bibnamefont {Lu}}, \bibinfo {author} {\bibfnamefont
  {P.}~\bibnamefont {David}}, \bibinfo {author} {\bibfnamefont
  {D.}~\bibnamefont {Dong}}, \bibinfo {author} {\bibfnamefont {H.}~\bibnamefont
  {Chen}}, \bibinfo {author} {\bibfnamefont {A.~S.}\ \bibnamefont {Etman}},
  \bibinfo {author} {\bibfnamefont {T.}~\bibnamefont {Lei}}, \bibinfo {author}
  {\bibfnamefont {J.}~\bibnamefont {Sun}}, \bibinfo {author} {\bibfnamefont
  {M.}~\bibnamefont {Grünwald}}, \bibinfo {author} {\bibfnamefont {D.~T.}\
  \bibnamefont {Limmer}}, \bibinfo {author} {\bibfnamefont {P.}~\bibnamefont
  {Yang}}, \ and\ \bibinfo {author} {\bibfnamefont {N.~S.}\ \bibnamefont
  {Ginsberg}},\ }\href {\doibase 10.1016/j.matt.2020.07.015} {\bibfield
  {journal} {\bibinfo  {journal} {Matter}\ }\textbf {\bibinfo {volume} {3}},\
  \bibinfo {pages} {534} (\bibinfo {year} {2020})},\ \Eprint
  {http://arxiv.org/abs/1907.13509} {1907.13509} \BibitemShut {NoStop}%
\bibitem [{\citenamefont {Gao}\ \emph {et~al.}(2023)\citenamefont {Gao},
  \citenamefont {Park}, \citenamefont {Jin}, \citenamefont {Chen},
  \citenamefont {Devyldere}, \citenamefont {Yang}, \citenamefont {Song},
  \citenamefont {Lin}, \citenamefont {Zhao}, \citenamefont {Siron} \emph
  {et~al.}}]{gao2023direct}%
  \BibitemOpen
  \bibfield  {author} {\bibinfo {author} {\bibfnamefont {M.}~\bibnamefont
  {Gao}}, \bibinfo {author} {\bibfnamefont {Y.}~\bibnamefont {Park}}, \bibinfo
  {author} {\bibfnamefont {J.}~\bibnamefont {Jin}}, \bibinfo {author}
  {\bibfnamefont {P.-C.}\ \bibnamefont {Chen}}, \bibinfo {author}
  {\bibfnamefont {H.}~\bibnamefont {Devyldere}}, \bibinfo {author}
  {\bibfnamefont {Y.}~\bibnamefont {Yang}}, \bibinfo {author} {\bibfnamefont
  {C.}~\bibnamefont {Song}}, \bibinfo {author} {\bibfnamefont {Z.}~\bibnamefont
  {Lin}}, \bibinfo {author} {\bibfnamefont {Q.}~\bibnamefont {Zhao}}, \bibinfo
  {author} {\bibfnamefont {M.}~\bibnamefont {Siron}},  \emph {et~al.},\
  }\href@noop {} {\bibfield  {journal} {\bibinfo  {journal} {Journal of the
  American Chemical Society}\ }\textbf {\bibinfo {volume} {145}},\ \bibinfo
  {pages} {4800} (\bibinfo {year} {2023})}\BibitemShut {NoStop}%
\bibitem [{\citenamefont {Quan}\ \emph {et~al.}(2021)\citenamefont {Quan},
  \citenamefont {Park}, \citenamefont {Guo}, \citenamefont {Gao}, \citenamefont
  {Jin}, \citenamefont {Huang}, \citenamefont {Copper}, \citenamefont
  {Schwartzberg}, \citenamefont {Schaller}, \citenamefont {Limmer} \emph
  {et~al.}}]{quan2021vibrational}%
  \BibitemOpen
  \bibfield  {author} {\bibinfo {author} {\bibfnamefont {L.~N.}\ \bibnamefont
  {Quan}}, \bibinfo {author} {\bibfnamefont {Y.}~\bibnamefont {Park}}, \bibinfo
  {author} {\bibfnamefont {P.}~\bibnamefont {Guo}}, \bibinfo {author}
  {\bibfnamefont {M.}~\bibnamefont {Gao}}, \bibinfo {author} {\bibfnamefont
  {J.}~\bibnamefont {Jin}}, \bibinfo {author} {\bibfnamefont {J.}~\bibnamefont
  {Huang}}, \bibinfo {author} {\bibfnamefont {J.~K.}\ \bibnamefont {Copper}},
  \bibinfo {author} {\bibfnamefont {A.}~\bibnamefont {Schwartzberg}}, \bibinfo
  {author} {\bibfnamefont {R.}~\bibnamefont {Schaller}}, \bibinfo {author}
  {\bibfnamefont {D.~T.}\ \bibnamefont {Limmer}},  \emph {et~al.},\ }\href@noop
  {} {\bibfield  {journal} {\bibinfo  {journal} {Proceedings of the National
  Academy of Sciences}\ }\textbf {\bibinfo {volume} {118}},\ \bibinfo {pages}
  {e2104425118} (\bibinfo {year} {2021})}\BibitemShut {NoStop}%
\bibitem [{\citenamefont {Sajedi}\ \emph {et~al.}(2022)\citenamefont {Sajedi},
  \citenamefont {Krivenkov}, \citenamefont {Marchenko}, \citenamefont
  {S{\'a}nchez-Barriga}, \citenamefont {Chandran}, \citenamefont {Varykhalov},
  \citenamefont {Rienks}, \citenamefont {Aguilera}, \citenamefont
  {Bl{\"u}gel},\ and\ \citenamefont {Rader}}]{sajedi2022there}%
  \BibitemOpen
  \bibfield  {author} {\bibinfo {author} {\bibfnamefont {M.}~\bibnamefont
  {Sajedi}}, \bibinfo {author} {\bibfnamefont {M.}~\bibnamefont {Krivenkov}},
  \bibinfo {author} {\bibfnamefont {D.}~\bibnamefont {Marchenko}}, \bibinfo
  {author} {\bibfnamefont {J.}~\bibnamefont {S{\'a}nchez-Barriga}}, \bibinfo
  {author} {\bibfnamefont {A.~K.}\ \bibnamefont {Chandran}}, \bibinfo {author}
  {\bibfnamefont {A.}~\bibnamefont {Varykhalov}}, \bibinfo {author}
  {\bibfnamefont {E.~D.}\ \bibnamefont {Rienks}}, \bibinfo {author}
  {\bibfnamefont {I.}~\bibnamefont {Aguilera}}, \bibinfo {author}
  {\bibfnamefont {S.}~\bibnamefont {Bl{\"u}gel}}, \ and\ \bibinfo {author}
  {\bibfnamefont {O.}~\bibnamefont {Rader}},\ }\href@noop {} {\bibfield
  {journal} {\bibinfo  {journal} {Physical Review Letters}\ }\textbf {\bibinfo
  {volume} {128}},\ \bibinfo {pages} {176405} (\bibinfo {year}
  {2022})}\BibitemShut {NoStop}%
\bibitem [{\citenamefont {Miyata}\ \emph {et~al.}(2017)\citenamefont {Miyata},
  \citenamefont {Meggiolaro}, \citenamefont {Trinh}, \citenamefont {Joshi},
  \citenamefont {Mosconi}, \citenamefont {Jones}, \citenamefont {Angelis},\
  and\ \citenamefont {Zhu}}]{sciadvzhu}%
  \BibitemOpen
  \bibfield  {author} {\bibinfo {author} {\bibfnamefont {K.}~\bibnamefont
  {Miyata}}, \bibinfo {author} {\bibfnamefont {D.}~\bibnamefont {Meggiolaro}},
  \bibinfo {author} {\bibfnamefont {M.~T.}\ \bibnamefont {Trinh}}, \bibinfo
  {author} {\bibfnamefont {P.~P.}\ \bibnamefont {Joshi}}, \bibinfo {author}
  {\bibfnamefont {E.}~\bibnamefont {Mosconi}}, \bibinfo {author} {\bibfnamefont
  {S.~C.}\ \bibnamefont {Jones}}, \bibinfo {author} {\bibfnamefont {F.~D.}\
  \bibnamefont {Angelis}}, \ and\ \bibinfo {author} {\bibfnamefont {X.-Y.}\
  \bibnamefont {Zhu}},\ }\href {\doibase 10.1126/sciadv.1701217} {\bibfield
  {journal} {\bibinfo  {journal} {Science Advances}\ }\textbf {\bibinfo
  {volume} {3}},\ \bibinfo {pages} {e1701217} (\bibinfo {year}
  {2017})}\BibitemShut {NoStop}%
\bibitem [{\citenamefont {Ceperley}(1995)}]{ceperley1995path}%
  \BibitemOpen
  \bibfield  {author} {\bibinfo {author} {\bibfnamefont {D.~M.}\ \bibnamefont
  {Ceperley}},\ }\href@noop {} {\bibfield  {journal} {\bibinfo  {journal}
  {Reviews of Modern Physics}\ }\textbf {\bibinfo {volume} {67}},\ \bibinfo
  {pages} {279} (\bibinfo {year} {1995})}\BibitemShut {NoStop}%
\bibitem [{\citenamefont {Schnitker}\ and\ \citenamefont
  {Rossky}(1987)}]{schnitker1987electron}%
  \BibitemOpen
  \bibfield  {author} {\bibinfo {author} {\bibfnamefont {J.}~\bibnamefont
  {Schnitker}}\ and\ \bibinfo {author} {\bibfnamefont {P.~J.}\ \bibnamefont
  {Rossky}},\ }\href@noop {} {\bibfield  {journal} {\bibinfo  {journal} {The
  Journal of Chemical Physics}\ }\textbf {\bibinfo {volume} {86}},\ \bibinfo
  {pages} {3462} (\bibinfo {year} {1987})}\BibitemShut {NoStop}%
\bibitem [{\citenamefont {Kuharski}\ \emph {et~al.}(1988)\citenamefont
  {Kuharski}, \citenamefont {Bader}, \citenamefont {Chandler}, \citenamefont
  {Sprik}, \citenamefont {Klein},\ and\ \citenamefont
  {Impey}}]{kuharski1988molecular}%
  \BibitemOpen
  \bibfield  {author} {\bibinfo {author} {\bibfnamefont {R.~A.}\ \bibnamefont
  {Kuharski}}, \bibinfo {author} {\bibfnamefont {J.~S.}\ \bibnamefont {Bader}},
  \bibinfo {author} {\bibfnamefont {D.}~\bibnamefont {Chandler}}, \bibinfo
  {author} {\bibfnamefont {M.}~\bibnamefont {Sprik}}, \bibinfo {author}
  {\bibfnamefont {M.~L.}\ \bibnamefont {Klein}}, \ and\ \bibinfo {author}
  {\bibfnamefont {R.~W.}\ \bibnamefont {Impey}},\ }\href@noop {} {\bibfield
  {journal} {\bibinfo  {journal} {The Journal of Chemical Physics}\ }\textbf
  {\bibinfo {volume} {89}},\ \bibinfo {pages} {3248} (\bibinfo {year}
  {1988})}\BibitemShut {NoStop}%
\bibitem [{\citenamefont {Wang}\ and\ \citenamefont
  {Zunger}(1994)}]{wang1994dielectric}%
  \BibitemOpen
  \bibfield  {author} {\bibinfo {author} {\bibfnamefont {L.-W.}\ \bibnamefont
  {Wang}}\ and\ \bibinfo {author} {\bibfnamefont {A.}~\bibnamefont {Zunger}},\
  }\href@noop {} {\bibfield  {journal} {\bibinfo  {journal} {Physical Review
  Letters}\ }\textbf {\bibinfo {volume} {73}},\ \bibinfo {pages} {1039}
  (\bibinfo {year} {1994})}\BibitemShut {NoStop}%
\bibitem [{\citenamefont {Feynman}(1998)}]{feynman1}%
  \BibitemOpen
  \bibfield  {author} {\bibinfo {author} {\bibfnamefont {R.~P.}\ \bibnamefont
  {Feynman}},\ }\href@noop {} {\emph {\bibinfo {title} {{Statistical
  Mechanics}}}}\ (\bibinfo  {publisher} {Westview},\ \bibinfo {year}
  {1998})\BibitemShut {NoStop}%
\bibitem [{\citenamefont {Feynman}\ and\ \citenamefont
  {Hibbs}(2005)}]{feynman2}%
  \BibitemOpen
  \bibfield  {author} {\bibinfo {author} {\bibfnamefont {R.~P.}\ \bibnamefont
  {Feynman}}\ and\ \bibinfo {author} {\bibfnamefont {A.~R.}\ \bibnamefont
  {Hibbs}},\ }\href@noop {} {\emph {\bibinfo {title} {{Quantum Mechanics and
  Path Integrals}}}}\ (\bibinfo  {publisher} {Dover},\ \bibinfo {year}
  {2005})\BibitemShut {NoStop}%
\bibitem [{\citenamefont {Habershon}\ \emph {et~al.}(2013)\citenamefont
  {Habershon}, \citenamefont {Manolopoulos}, \citenamefont {Markland},\ and\
  \citenamefont {Miller~III}}]{habershon2013ring}%
  \BibitemOpen
  \bibfield  {author} {\bibinfo {author} {\bibfnamefont {S.}~\bibnamefont
  {Habershon}}, \bibinfo {author} {\bibfnamefont {D.~E.}\ \bibnamefont
  {Manolopoulos}}, \bibinfo {author} {\bibfnamefont {T.~E.}\ \bibnamefont
  {Markland}}, \ and\ \bibinfo {author} {\bibfnamefont {T.~F.}\ \bibnamefont
  {Miller~III}},\ }\href@noop {} {\bibfield  {journal} {\bibinfo  {journal}
  {Annual review of physical chemistry}\ }\textbf {\bibinfo {volume} {64}},\
  \bibinfo {pages} {387} (\bibinfo {year} {2013})}\BibitemShut {NoStop}%
\bibitem [{\citenamefont {Plimpton}(1995)}]{lammps}%
  \BibitemOpen
  \bibfield  {author} {\bibinfo {author} {\bibfnamefont {S.}~\bibnamefont
  {Plimpton}},\ }\href@noop {} {\bibfield  {journal} {\bibinfo  {journal} {J
  Comp Phys}\ }\textbf {\bibinfo {volume} {117}},\ \bibinfo {pages} {1}
  (\bibinfo {year} {1995})}\BibitemShut {NoStop}%
\bibitem [{\citenamefont {Scharf}\ \emph {et~al.}(1986)\citenamefont {Scharf},
  \citenamefont {Jortner},\ and\ \citenamefont {Landman}}]{binde}%
  \BibitemOpen
  \bibfield  {author} {\bibinfo {author} {\bibfnamefont {D.}~\bibnamefont
  {Scharf}}, \bibinfo {author} {\bibfnamefont {J.}~\bibnamefont {Jortner}}, \
  and\ \bibinfo {author} {\bibfnamefont {U.}~\bibnamefont {Landman}},\ }\href
  {\doibase 10.1016/0009-2614(86)80247-0} {\bibfield  {journal} {\bibinfo
  {journal} {Chemical Physics Letters}\ }\textbf {\bibinfo {volume} {130}},\
  \bibinfo {pages} {504} (\bibinfo {year} {1986})}\BibitemShut {NoStop}%
\bibitem [{\citenamefont {Herman}\ \emph {et~al.}(1982)\citenamefont {Herman},
  \citenamefont {Bruskin},\ and\ \citenamefont {Berne}}]{virial}%
  \BibitemOpen
  \bibfield  {author} {\bibinfo {author} {\bibfnamefont {M.~F.}\ \bibnamefont
  {Herman}}, \bibinfo {author} {\bibfnamefont {E.~J.}\ \bibnamefont {Bruskin}},
  \ and\ \bibinfo {author} {\bibfnamefont {B.~J.}\ \bibnamefont {Berne}},\
  }\href {\doibase 10.1063/1.442815} {\bibfield  {journal} {\bibinfo  {journal}
  {The Journal of Chemical Physics}\ }\textbf {\bibinfo {volume} {76}},\
  \bibinfo {pages} {5150} (\bibinfo {year} {1982})}\BibitemShut {NoStop}%
\bibitem [{\citenamefont {Brennan}\ \emph {et~al.}(2020)\citenamefont
  {Brennan}, \citenamefont {Forde}, \citenamefont {Zhukovskyi}, \citenamefont
  {Baublis}, \citenamefont {Morozov}, \citenamefont {Zhang}, \citenamefont
  {Zhang}, \citenamefont {Kilin},\ and\ \citenamefont
  {Kuno}}]{10.1021/acs.jpclett.0c01407}%
  \BibitemOpen
  \bibfield  {author} {\bibinfo {author} {\bibfnamefont {M.~C.}\ \bibnamefont
  {Brennan}}, \bibinfo {author} {\bibfnamefont {A.}~\bibnamefont {Forde}},
  \bibinfo {author} {\bibfnamefont {M.}~\bibnamefont {Zhukovskyi}}, \bibinfo
  {author} {\bibfnamefont {A.~J.}\ \bibnamefont {Baublis}}, \bibinfo {author}
  {\bibfnamefont {Y.~V.}\ \bibnamefont {Morozov}}, \bibinfo {author}
  {\bibfnamefont {S.}~\bibnamefont {Zhang}}, \bibinfo {author} {\bibfnamefont
  {Z.}~\bibnamefont {Zhang}}, \bibinfo {author} {\bibfnamefont {D.~S.}\
  \bibnamefont {Kilin}}, \ and\ \bibinfo {author} {\bibfnamefont
  {M.}~\bibnamefont {Kuno}},\ }\href {\doibase 10.1021/acs.jpclett.0c01407}
  {\bibfield  {journal} {\bibinfo  {journal} {The Journal of Physical Chemistry
  Letters}\ }\textbf {\bibinfo {volume} {11}},\ \bibinfo {pages} {4937}
  (\bibinfo {year} {2020})}\BibitemShut {NoStop}%
\bibitem [{\citenamefont {Protesescu}\ \emph
  {et~al.}(2015{\natexlab{b}})\citenamefont {Protesescu}, \citenamefont
  {Yakunin}, \citenamefont {Bodnarchuk}, \citenamefont {Krieg}, \citenamefont
  {Caputo}, \citenamefont {Hendon}, \citenamefont {Yang}, \citenamefont
  {Walsh},\ and\ \citenamefont {Kovalenko}}]{protesescu2015nanocrystals}%
  \BibitemOpen
  \bibfield  {author} {\bibinfo {author} {\bibfnamefont {L.}~\bibnamefont
  {Protesescu}}, \bibinfo {author} {\bibfnamefont {S.}~\bibnamefont {Yakunin}},
  \bibinfo {author} {\bibfnamefont {M.~I.}\ \bibnamefont {Bodnarchuk}},
  \bibinfo {author} {\bibfnamefont {F.}~\bibnamefont {Krieg}}, \bibinfo
  {author} {\bibfnamefont {R.}~\bibnamefont {Caputo}}, \bibinfo {author}
  {\bibfnamefont {C.~H.}\ \bibnamefont {Hendon}}, \bibinfo {author}
  {\bibfnamefont {R.~X.}\ \bibnamefont {Yang}}, \bibinfo {author}
  {\bibfnamefont {A.}~\bibnamefont {Walsh}}, \ and\ \bibinfo {author}
  {\bibfnamefont {M.~V.}\ \bibnamefont {Kovalenko}},\ }\href@noop {} {\bibfield
   {journal} {\bibinfo  {journal} {Nano letters}\ }\textbf {\bibinfo {volume}
  {15}},\ \bibinfo {pages} {3692} (\bibinfo {year}
  {2015}{\natexlab{b}})}\BibitemShut {NoStop}%
\bibitem [{\citenamefont {Park}\ \emph {et~al.}(2000)\citenamefont {Park},
  \citenamefont {Jeen}, \citenamefont {Kim},\ and\ \citenamefont
  {Kim}}]{biexfitting}%
  \BibitemOpen
  \bibfield  {author} {\bibinfo {author} {\bibfnamefont {S.}~\bibnamefont
  {Park}}, \bibinfo {author} {\bibfnamefont {G.}~\bibnamefont {Jeen}}, \bibinfo
  {author} {\bibfnamefont {H.}~\bibnamefont {Kim}}, \ and\ \bibinfo {author}
  {\bibfnamefont {I.}~\bibnamefont {Kim}},\ }\href@noop {} {\bibfield
  {journal} {\bibinfo  {journal} {Journal of the Korean Physical Society}\
  }\textbf {\bibinfo {volume} {37}},\ \bibinfo {pages} {309} (\bibinfo {year}
  {2000})}\BibitemShut {NoStop}%
\bibitem [{\citenamefont {Hu}\ \emph {et~al.}(1990)\citenamefont {Hu},
  \citenamefont {Koch}, \citenamefont {Lindberg}, \citenamefont
  {Peyghambarian}, \citenamefont {Pollock},\ and\ \citenamefont
  {Abraham}}]{10.1103/physrevlett.64.1805}%
  \BibitemOpen
  \bibfield  {author} {\bibinfo {author} {\bibfnamefont {Y.~Z.}\ \bibnamefont
  {Hu}}, \bibinfo {author} {\bibfnamefont {S.~W.}\ \bibnamefont {Koch}},
  \bibinfo {author} {\bibfnamefont {M.}~\bibnamefont {Lindberg}}, \bibinfo
  {author} {\bibfnamefont {N.}~\bibnamefont {Peyghambarian}}, \bibinfo {author}
  {\bibfnamefont {E.~L.}\ \bibnamefont {Pollock}}, \ and\ \bibinfo {author}
  {\bibfnamefont {F.~F.}\ \bibnamefont {Abraham}},\ }\href {\doibase
  10.1103/physrevlett.64.1805} {\bibfield  {journal} {\bibinfo  {journal}
  {Physical Review Letters}\ }\textbf {\bibinfo {volume} {64}},\ \bibinfo
  {pages} {1805} (\bibinfo {year} {1990})}\BibitemShut {NoStop}%
\bibitem [{\citenamefont {Guo}\ \emph {et~al.}(2019)\citenamefont {Guo},
  \citenamefont {Yaffe}, \citenamefont {Hull}, \citenamefont {Owen},
  \citenamefont {Reichman},\ and\ \citenamefont
  {Brus}}]{10.1038/s41467-019-09057-5}%
  \BibitemOpen
  \bibfield  {author} {\bibinfo {author} {\bibfnamefont {Y.}~\bibnamefont
  {Guo}}, \bibinfo {author} {\bibfnamefont {O.}~\bibnamefont {Yaffe}}, \bibinfo
  {author} {\bibfnamefont {T.~D.}\ \bibnamefont {Hull}}, \bibinfo {author}
  {\bibfnamefont {J.~S.}\ \bibnamefont {Owen}}, \bibinfo {author}
  {\bibfnamefont {D.~R.}\ \bibnamefont {Reichman}}, \ and\ \bibinfo {author}
  {\bibfnamefont {L.~E.}\ \bibnamefont {Brus}},\ }\href {\doibase
  10.1038/s41467-019-09057-5} {\bibfield  {journal} {\bibinfo  {journal}
  {Nature Communications}\ }\textbf {\bibinfo {volume} {10}},\ \bibinfo {pages}
  {1175} (\bibinfo {year} {2019})}\BibitemShut {NoStop}%
\bibitem [{\citenamefont {Brennan}\ \emph {et~al.}(2017)\citenamefont
  {Brennan}, \citenamefont {Herr}, \citenamefont {Nguyen-Beck}, \citenamefont
  {Zinna}, \citenamefont {Draguta}, \citenamefont {Rouvimov}, \citenamefont
  {Parkhill},\ and\ \citenamefont {Kuno}}]{10.1021/jacs.7b05683}%
  \BibitemOpen
  \bibfield  {author} {\bibinfo {author} {\bibfnamefont {M.~C.}\ \bibnamefont
  {Brennan}}, \bibinfo {author} {\bibfnamefont {J.~E.}\ \bibnamefont {Herr}},
  \bibinfo {author} {\bibfnamefont {T.~S.}\ \bibnamefont {Nguyen-Beck}},
  \bibinfo {author} {\bibfnamefont {J.}~\bibnamefont {Zinna}}, \bibinfo
  {author} {\bibfnamefont {S.}~\bibnamefont {Draguta}}, \bibinfo {author}
  {\bibfnamefont {S.}~\bibnamefont {Rouvimov}}, \bibinfo {author}
  {\bibfnamefont {J.}~\bibnamefont {Parkhill}}, \ and\ \bibinfo {author}
  {\bibfnamefont {M.}~\bibnamefont {Kuno}},\ }\href {\doibase
  10.1021/jacs.7b05683} {\bibfield  {journal} {\bibinfo  {journal} {Journal of
  the American Chemical Society}\ }\textbf {\bibinfo {volume} {139}},\ \bibinfo
  {pages} {12201} (\bibinfo {year} {2017})}\BibitemShut {NoStop}%
\end{thebibliography}%


\begin{thebibliography}{12}%
\makeatletter
\providecommand \@ifxundefined [1]{%
 \@ifx{#1\undefined}
}%
\providecommand \@ifnum [1]{%
 \ifnum #1\expandafter \@firstoftwo
 \else \expandafter \@secondoftwo
 \fi
}%
\providecommand \@ifx [1]{%
 \ifx #1\expandafter \@firstoftwo
 \else \expandafter \@secondoftwo
 \fi
}%
\providecommand \natexlab [1]{#1}%
\providecommand \enquote  [1]{``#1''}%
\providecommand \bibnamefont  [1]{#1}%
\providecommand \bibfnamefont [1]{#1}%
\providecommand \citenamefont [1]{#1}%
\providecommand \href@noop [0]{\@secondoftwo}%
\providecommand \href [0]{\begingroup \@sanitize@url \@href}%
\providecommand \@href[1]{\@@startlink{#1}\@@href}%
\providecommand \@@href[1]{\endgroup#1\@@endlink}%
\providecommand \@sanitize@url [0]{\catcode `\\12\catcode `\$12\catcode
  `\&12\catcode `\#12\catcode `\^12\catcode `\_12\catcode `\%12\relax}%
\providecommand \@@startlink[1]{}%
\providecommand \@@endlink[0]{}%
\providecommand \url  [0]{\begingroup\@sanitize@url \@url }%
\providecommand \@url [1]{\endgroup\@href {#1}{\urlprefix }}%
\providecommand \urlprefix  [0]{URL }%
\providecommand \Eprint [0]{\href }%
\providecommand \doibase [0]{http://dx.doi.org/}%
\providecommand \selectlanguage [0]{\@gobble}%
\providecommand \bibinfo  [0]{\@secondoftwo}%
\providecommand \bibfield  [0]{\@secondoftwo}%
\providecommand \translation [1]{[#1]}%
\providecommand \BibitemOpen [0]{}%
\providecommand \bibitemStop [0]{}%
\providecommand \bibitemNoStop [0]{.\EOS\space}%
\providecommand \EOS [0]{\spacefactor3000\relax}%
\providecommand \BibitemShut  [1]{\csname bibitem#1\endcsname}%
\let\auto@bib@innerbib\@empty
\bibitem [{\citenamefont {Weinberg}\ \emph {et~al.}(2023)\citenamefont
  {Weinberg}, \citenamefont {Park}, \citenamefont {Limmer},\ and\ \citenamefont
  {Rabani}}]{danielarxiv}%
  \BibitemOpen
  \bibfield  {author} {\bibinfo {author} {\bibfnamefont {D.}~\bibnamefont
  {Weinberg}}, \bibinfo {author} {\bibfnamefont {Y.}~\bibnamefont {Park}},
  \bibinfo {author} {\bibfnamefont {D.~T.}\ \bibnamefont {Limmer}}, \ and\
  \bibinfo {author} {\bibfnamefont {E.}~\bibnamefont {Rabani}},\ }\href@noop {}
  {\bibfield  {journal} {\bibinfo  {journal} {Nano Letters}\ } (\bibinfo {year}
  {2023})}\BibitemShut {NoStop}%
\bibitem [{\citenamefont {Bischak}\ \emph {et~al.}(2020)\citenamefont
  {Bischak}, \citenamefont {Lai}, \citenamefont {Fan}, \citenamefont {Lu},
  \citenamefont {David}, \citenamefont {Dong}, \citenamefont {Chen},
  \citenamefont {Etman}, \citenamefont {Lei}, \citenamefont {Sun},
  \citenamefont {Grünwald}, \citenamefont {Limmer}, \citenamefont {Yang},\
  and\ \citenamefont {Ginsberg}}]{10.1016/j.matt.2020.07.015}%
  \BibitemOpen
  \bibfield  {author} {\bibinfo {author} {\bibfnamefont {C.~G.}\ \bibnamefont
  {Bischak}}, \bibinfo {author} {\bibfnamefont {M.}~\bibnamefont {Lai}},
  \bibinfo {author} {\bibfnamefont {Z.}~\bibnamefont {Fan}}, \bibinfo {author}
  {\bibfnamefont {D.}~\bibnamefont {Lu}}, \bibinfo {author} {\bibfnamefont
  {P.}~\bibnamefont {David}}, \bibinfo {author} {\bibfnamefont
  {D.}~\bibnamefont {Dong}}, \bibinfo {author} {\bibfnamefont {H.}~\bibnamefont
  {Chen}}, \bibinfo {author} {\bibfnamefont {A.~S.}\ \bibnamefont {Etman}},
  \bibinfo {author} {\bibfnamefont {T.}~\bibnamefont {Lei}}, \bibinfo {author}
  {\bibfnamefont {J.}~\bibnamefont {Sun}}, \bibinfo {author} {\bibfnamefont
  {M.}~\bibnamefont {Grünwald}}, \bibinfo {author} {\bibfnamefont {D.~T.}\
  \bibnamefont {Limmer}}, \bibinfo {author} {\bibfnamefont {P.}~\bibnamefont
  {Yang}}, \ and\ \bibinfo {author} {\bibfnamefont {N.~S.}\ \bibnamefont
  {Ginsberg}},\ }\href {\doibase 10.1016/j.matt.2020.07.015} {\bibfield
  {journal} {\bibinfo  {journal} {Matter}\ }\textbf {\bibinfo {volume} {3}},\
  \bibinfo {pages} {534} (\bibinfo {year} {2020})},\ \Eprint
  {http://arxiv.org/abs/1907.13509} {1907.13509} \BibitemShut {NoStop}%
\bibitem [{\citenamefont {Zhao}\ \emph {et~al.}(2020)\citenamefont {Zhao},
  \citenamefont {Hazarika}, \citenamefont {Schelhas}, \citenamefont {Liu},
  \citenamefont {Gaulding}, \citenamefont {Li}, \citenamefont {Zhang},
  \citenamefont {Toney}, \citenamefont {Sercel},\ and\ \citenamefont
  {Luther}}]{10.1021/acsenergylett.9b02395}%
  \BibitemOpen
  \bibfield  {author} {\bibinfo {author} {\bibfnamefont {Q.}~\bibnamefont
  {Zhao}}, \bibinfo {author} {\bibfnamefont {A.}~\bibnamefont {Hazarika}},
  \bibinfo {author} {\bibfnamefont {L.~T.}\ \bibnamefont {Schelhas}}, \bibinfo
  {author} {\bibfnamefont {J.}~\bibnamefont {Liu}}, \bibinfo {author}
  {\bibfnamefont {E.~A.}\ \bibnamefont {Gaulding}}, \bibinfo {author}
  {\bibfnamefont {G.}~\bibnamefont {Li}}, \bibinfo {author} {\bibfnamefont
  {M.}~\bibnamefont {Zhang}}, \bibinfo {author} {\bibfnamefont {M.~F.}\
  \bibnamefont {Toney}}, \bibinfo {author} {\bibfnamefont {P.~C.}\ \bibnamefont
  {Sercel}}, \ and\ \bibinfo {author} {\bibfnamefont {J.~M.}\ \bibnamefont
  {Luther}},\ }\href {\doibase 10.1021/acsenergylett.9b02395} {\bibfield
  {journal} {\bibinfo  {journal} {ACS Energy Letters}\ }\textbf {\bibinfo
  {volume} {5}},\ \bibinfo {pages} {238} (\bibinfo {year} {2020})}\BibitemShut
  {NoStop}%
\bibitem [{\citenamefont {Yang}\ and\ \citenamefont
  {Tan}(2020)}]{10.1063/1.5128016}%
  \BibitemOpen
  \bibfield  {author} {\bibinfo {author} {\bibfnamefont {R.~X.}\ \bibnamefont
  {Yang}}\ and\ \bibinfo {author} {\bibfnamefont {L.~Z.}\ \bibnamefont {Tan}},\
  }\href {\doibase 10.1063/1.5128016} {\bibfield  {journal} {\bibinfo
  {journal} {The Journal of Chemical Physics}\ }\textbf {\bibinfo {volume}
  {152}},\ \bibinfo {pages} {034702} (\bibinfo {year} {2020})}\BibitemShut
  {NoStop}%
\bibitem [{\citenamefont {Yang}\ \emph {et~al.}(2017)\citenamefont {Yang},
  \citenamefont {Skelton}, \citenamefont {Silva}, \citenamefont {Frost},\ and\
  \citenamefont {Walsh}}]{acs.jpclett.7b02423}%
  \BibitemOpen
  \bibfield  {author} {\bibinfo {author} {\bibfnamefont {R.~X.}\ \bibnamefont
  {Yang}}, \bibinfo {author} {\bibfnamefont {J.~M.}\ \bibnamefont {Skelton}},
  \bibinfo {author} {\bibfnamefont {E.~L.~d.}\ \bibnamefont {Silva}}, \bibinfo
  {author} {\bibfnamefont {J.~M.}\ \bibnamefont {Frost}}, \ and\ \bibinfo
  {author} {\bibfnamefont {A.}~\bibnamefont {Walsh}},\ }\href {\doibase
  10.1021/acs.jpclett.7b02423} {\bibfield  {journal} {\bibinfo  {journal} {The
  Journal of Physical Chemistry Letters}\ }\textbf {\bibinfo {volume} {8}},\
  \bibinfo {pages} {4720} (\bibinfo {year} {2017})},\ \Eprint
  {http://arxiv.org/abs/1708.00499} {1708.00499} \BibitemShut {NoStop}%
\bibitem [{\citenamefont {Han}\ \emph {et~al.}(2022)\citenamefont {Han},
  \citenamefont {Liang}, \citenamefont {Lin}, \citenamefont {Li}, \citenamefont
  {Sun}, \citenamefont {Zhang}, \citenamefont {Sercel},\ and\ \citenamefont
  {Wu}}]{10.1038/s41563-022-01349-4}%
  \BibitemOpen
  \bibfield  {author} {\bibinfo {author} {\bibfnamefont {Y.}~\bibnamefont
  {Han}}, \bibinfo {author} {\bibfnamefont {W.}~\bibnamefont {Liang}}, \bibinfo
  {author} {\bibfnamefont {X.}~\bibnamefont {Lin}}, \bibinfo {author}
  {\bibfnamefont {Y.}~\bibnamefont {Li}}, \bibinfo {author} {\bibfnamefont
  {F.}~\bibnamefont {Sun}}, \bibinfo {author} {\bibfnamefont {F.}~\bibnamefont
  {Zhang}}, \bibinfo {author} {\bibfnamefont {P.~C.}\ \bibnamefont {Sercel}}, \
  and\ \bibinfo {author} {\bibfnamefont {K.}~\bibnamefont {Wu}},\ }\href
  {\doibase 10.1038/s41563-022-01349-4} {\bibfield  {journal} {\bibinfo
  {journal} {Nature Materials}\ }\textbf {\bibinfo {volume} {21}},\ \bibinfo
  {pages} {1282} (\bibinfo {year} {2022})},\ \Eprint
  {http://arxiv.org/abs/2206.13716} {2206.13716} \BibitemShut {NoStop}%
\bibitem [{\citenamefont {Miyata}\ \emph {et~al.}(2017)\citenamefont {Miyata},
  \citenamefont {Meggiolaro}, \citenamefont {Trinh}, \citenamefont {Joshi},
  \citenamefont {Mosconi}, \citenamefont {Jones}, \citenamefont {Angelis},\
  and\ \citenamefont {Zhu}}]{sciadvzhu}%
  \BibitemOpen
  \bibfield  {author} {\bibinfo {author} {\bibfnamefont {K.}~\bibnamefont
  {Miyata}}, \bibinfo {author} {\bibfnamefont {D.}~\bibnamefont {Meggiolaro}},
  \bibinfo {author} {\bibfnamefont {M.~T.}\ \bibnamefont {Trinh}}, \bibinfo
  {author} {\bibfnamefont {P.~P.}\ \bibnamefont {Joshi}}, \bibinfo {author}
  {\bibfnamefont {E.}~\bibnamefont {Mosconi}}, \bibinfo {author} {\bibfnamefont
  {S.~C.}\ \bibnamefont {Jones}}, \bibinfo {author} {\bibfnamefont {F.~D.}\
  \bibnamefont {Angelis}}, \ and\ \bibinfo {author} {\bibfnamefont {X.-Y.}\
  \bibnamefont {Zhu}},\ }\href {\doibase 10.1126/sciadv.1701217} {\bibfield
  {journal} {\bibinfo  {journal} {Science Advances}\ }\textbf {\bibinfo
  {volume} {3}},\ \bibinfo {pages} {e1701217} (\bibinfo {year}
  {2017})}\BibitemShut {NoStop}%
\bibitem [{\citenamefont {Sometani}(2000)}]{imagemethod}%
  \BibitemOpen
  \bibfield  {author} {\bibinfo {author} {\bibfnamefont {T.}~\bibnamefont
  {Sometani}},\ }\href {\doibase 10.1088/0143-0807/21/6/305} {\bibfield
  {journal} {\bibinfo  {journal} {European Journal of Physics}\ }\textbf
  {\bibinfo {volume} {21}},\ \bibinfo {pages} {549} (\bibinfo {year}
  {2000})}\BibitemShut {NoStop}%
\bibitem [{\citenamefont {Plimpton}(1995)}]{lammps}%
  \BibitemOpen
  \bibfield  {author} {\bibinfo {author} {\bibfnamefont {S.}~\bibnamefont
  {Plimpton}},\ }\href@noop {} {\bibfield  {journal} {\bibinfo  {journal} {J
  Comp Phys}\ }\textbf {\bibinfo {volume} {117}},\ \bibinfo {pages} {1}
  (\bibinfo {year} {1995})}\BibitemShut {NoStop}%
\bibitem [{\citenamefont {Park}\ and\ \citenamefont
  {Limmer}(2022)}]{parkjcp2022}%
  \BibitemOpen
  \bibfield  {author} {\bibinfo {author} {\bibfnamefont {Y.}~\bibnamefont
  {Park}}\ and\ \bibinfo {author} {\bibfnamefont {D.~T.}\ \bibnamefont
  {Limmer}},\ }\href {\doibase 10.1063/5.0100738} {\bibfield  {journal}
  {\bibinfo  {journal} {The Journal of Chemical Physics}\ }\textbf {\bibinfo
  {volume} {157}},\ \bibinfo {pages} {104116} (\bibinfo {year} {2022})},\
  \Eprint {http://arxiv.org/abs/2205.11780} {2205.11780} \BibitemShut {NoStop}%
\bibitem [{\citenamefont {Scharf}\ \emph {et~al.}(1986)\citenamefont {Scharf},
  \citenamefont {Jortner},\ and\ \citenamefont {Landman}}]{binde}%
  \BibitemOpen
  \bibfield  {author} {\bibinfo {author} {\bibfnamefont {D.}~\bibnamefont
  {Scharf}}, \bibinfo {author} {\bibfnamefont {J.}~\bibnamefont {Jortner}}, \
  and\ \bibinfo {author} {\bibfnamefont {U.}~\bibnamefont {Landman}},\ }\href
  {\doibase 10.1016/0009-2614(86)80247-0} {\bibfield  {journal} {\bibinfo
  {journal} {Chemical Physics Letters}\ }\textbf {\bibinfo {volume} {130}},\
  \bibinfo {pages} {504} (\bibinfo {year} {1986})}\BibitemShut {NoStop}%
\bibitem [{\citenamefont {Herman}\ \emph {et~al.}(1982)\citenamefont {Herman},
  \citenamefont {Bruskin},\ and\ \citenamefont {Berne}}]{virial}%
  \BibitemOpen
  \bibfield  {author} {\bibinfo {author} {\bibfnamefont {M.~F.}\ \bibnamefont
  {Herman}}, \bibinfo {author} {\bibfnamefont {E.~J.}\ \bibnamefont {Bruskin}},
  \ and\ \bibinfo {author} {\bibfnamefont {B.~J.}\ \bibnamefont {Berne}},\
  }\href {\doibase 10.1063/1.442815} {\bibfield  {journal} {\bibinfo  {journal}
  {The Journal of Chemical Physics}\ }\textbf {\bibinfo {volume} {76}},\
  \bibinfo {pages} {5150} (\bibinfo {year} {1982})}\BibitemShut {NoStop}%
\end{thebibliography}%
\end{document}


\title{Supporting Information to ``Biexcitons are bound in CsPbBr$_3$ Perovskite Nanocrystals"}

\author{Yoonjae Park}
 \affiliation{Department of Chemistry, University of California, Berkeley}

\author{David T. Limmer}
 \email{dlimmer@berkeley.edu}
 \affiliation{Department of Chemistry, University of California, Berkeley}
\affiliation{Materials Science Division, Lawrence Berkeley National Laboratory}
\affiliation{Chemical Science Division, Lawrence Berkeley National Laboratory}
\affiliation{Kavli Energy NanoScience Institute, Berkeley, California, Berkeley}

\date{\today}

\maketitle


\section{Simulation details with explicit perovskite nanocrystals}

For the simulations of exciton and biexciton with explicit CsPbBr$_3$ perovskite nanocrystals, we consider the range of nanocrystal sizes from 2.4 nm to 6 nm. For the perovskite nanocrystals, we take the relaxed structure for each size of nanocrystals \cite{danielarxiv} and adopt an atomistic force field \cite{10.1016/j.matt.2020.07.015}, whose pair-wise interactions between atoms with type $i$ and $j$ are given by the sum of Coulomb potential and Lennard-Jones (LJ) potential as
\begin{equation}
U_{} (x_{ij}) = \frac{q_i q_j}{4\pi \varepsilon_{0} x_{ij}} + 4 \varepsilon_{ij} \left[ \left(\frac{\sigma_{ij}}{x_{ij}}\right)^{12} - \left(\frac{\sigma_{ij}}{x_{ij}}\right)^{6} \right]
\label{biex_LJ}
\end{equation}
where $x_{ij} = |\mathbf{x}_i - \mathbf{x}_j|$ is the distance between two atoms and the parameters $q_{i}$, $\varepsilon_{ij}$ and $\sigma_{ij}$ are summarized in Table \ref{biex_LJ}. For each size of nanocrystal, since the total charge of the lattice is not neutral, a different partial charge is assigned for surface Cs atoms to reduce the effect from surface boundaries of nanocrystals and to stabilize the nanocrystals structure. This discharging of the surface is a simple model of the passivating ligands used synthetically \cite{10.1021/acsenergylett.9b02395, 10.1063/1.5128016}, relying on the fact that the lattice distortion of lead halide perovskites is largely determined by the lead halide octahedra \cite{acs.jpclett.7b02423, 10.1038/s41563-022-01349-4}. The partial charge of surface Cs atoms is defined as 
\begin{equation}
q_{\mathrm{surf}} = - \frac{q_{\mathrm{Cs}} N_{\mathrm{Cs}} + q_{\mathrm{Br}} N_{\mathrm{Pb}} + q_{\mathrm{Br}} N_{\mathrm{Br}}}{N_{\mathrm{surf}}}
\label{biex_qsur}
\end{equation}
where $q_i$ and $N_i$ with $i \in \{ \mathrm{Cs}, \mathrm{Pb}, \mathrm{Br}, \mathrm{Cs}_{\mathrm{surf}}\}$ are the charge and the number of atoms with type $i$ in the lattice with the subscript surf used for surface Cs atoms. 

\begin {figure*}
\centering\includegraphics [width=12.5cm] {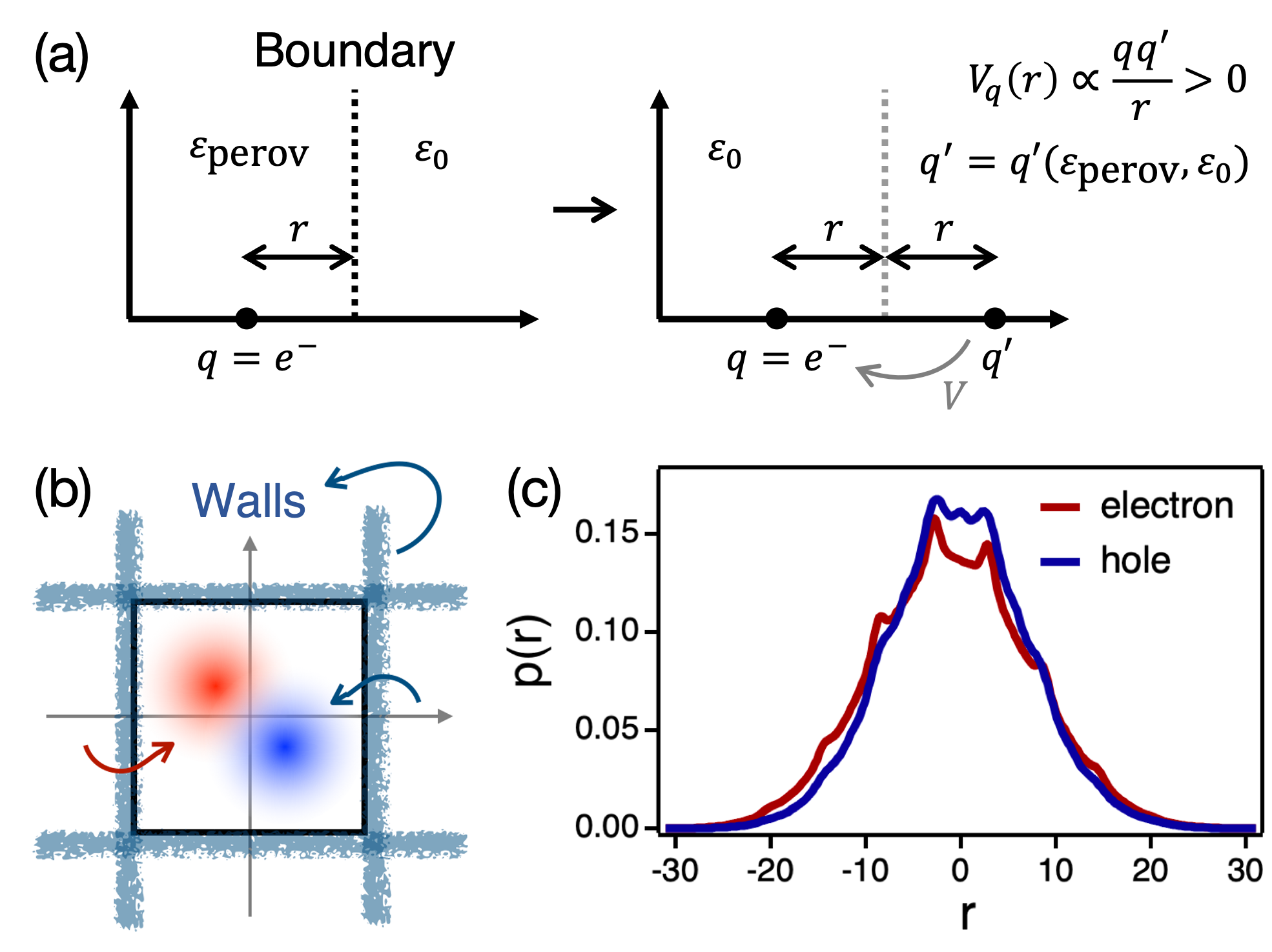}
\caption{(a) Schematics of how the effect of dielectric discontinuity on the charge $q$ inside the nanocrystal can be replaced by an image point charge where the potential $V_q(r)$ exerted on the inside charge $q$ which is $r$ distance away from the boundary due to the point charge $q'$ is determined by two different dielectric constants. (b) Schematics of nanocrystals with walls at the boundaries. (c) Charge density distributions of electron (red) and hole (blue) in the perovskite nanocrystal.}
\label{Fig_wall}
\end{figure*}

For electrons and holes quasiparticles, we use a path integral approach with a discretized Hamiltonian given by
\begin{equation}
\mathcal{H}_{\mathrm{el}} 
= \sum_{i} \sum_{t=1}^n \frac{m_i n}{2\beta^2 \hbar^2} ({\mathbf{x}}_{i,t} - {\mathbf{x}}_{i,t+1})^2
+ \sum_{i \ne j} \sum_{t=1}^n \frac{q_i q_j}{4\pi \varepsilon_{\infty} n|{\mathbf{x}}_{i,t} - {\mathbf{x}}_{j,t}|}
\label{biex_dHel}
\end{equation}
where $i,j \in \{e_1, h_1 \}$ or $i,j \in \{e_1, e_2, h_1, h_2 \}$ for a simulation with an exciton or biexciton, $n=1000$ is the number of discretization of the path integral, $\mathbf{x}_{i,t}$ is the position of $t^{th}$ bead in $i$ ring polymer, and $\mathbf{x}_{i,n+1} = \mathbf{x}_{i,1}$ for each quantum particle $i$. $m_i$ is the band mass which is set to 0.22\,$m_0$ and 0.24\,$m_0$ for electrons and holes \cite{sciadvzhu}, respectively, and $\varepsilon_{\infty} = 4.3$ is the optical dielectric constant in the unit of vacuum permittivity $\varepsilon_0$ \cite{sciadvzhu}. The last term of Eq.\,\ref{biex_dHel} can be also denoted as $\sum_{i \ne j} \mathcal{H}_{\mathrm{C}}^{ij}$. 
The discretized Hamiltonian for the interaction between charges and the lattice is described by the sum of pseudopotentials in the form of truncated Coulomb potential 
\begin{equation}
\begin{aligned}
\mathcal{H}_{\mathrm{int}} = \sum_i \mathcal{H}_{\mathrm{int}}^{i}
= \sum_{i} \sum_{t=1}^n  \sum_{j=1}^N \frac{q_i q_j}{4\pi \varepsilon_0 n \sqrt{ r_{\mathrm{cut}}^2 +|\mathbf{x}_{i,t}-\mathbf{x}_j|^2} }
\label{biex_dHint}
\end{aligned}
\end{equation}
where $1/|\mathbf{x}|$ is approximated by $(r_{\mathrm{cut}}^2 + \mathbf{x}^2)^{-1/2}$ as given in Eq.4. Considering that the similar band masses of electron and hole in perovskite imply the similar extent of delocalization of the charges, the cutoff parameters $r_{\mathrm{cut}}$ for each pair interaction between charges and different types of atoms are chosen to get the similar charge density distribution for electron and hole in the nanocrystals as shown in Fig.\ref{Fig_wall}\,(c), whose parameters are summarized in Table \ref{biex_pseudo}.
Additionally, since inside and outside of nanocrystals are different mediums, we implement wall potentials to take the resultant dielectric discontinuity into account. As schematically shown in Fig. \ref{Fig_wall} (a), the effect of dielectric discontinuity on the charge $q$ inside the nanocrystal can be replaced by an image point charge \cite{imagemethod}. The potential $V_q(r)$ exerted on the inside charge $q$, which is $r$ distance away from the boundary, due to the point charge $q'$ is determined by two different dielectric constants up to the first order as $V_q(r) = qq'/8\pi \varepsilon_0 r$ with $q' = q (\varepsilon_{\mathrm{perov}} - \varepsilon_0)/(\varepsilon_{\mathrm{perov}} + \varepsilon_0)$ and $\varepsilon_{\mathrm{perov}} = \varepsilon_{\infty}$, resulting in Eq.5. Since the resultant value of the potential is positive, this effective wall potential results in a dielectric confinement where the charges (ring polymers) stay inside the nanocrystals through the repulsive Coulomb interaction with walls at the boundaries as schematically shown in Fig. \ref{Fig_wall} (b). 

For the simulations of exciton and biexciton, ring polymers are equilibrated for 500\,ps in NVT ensemble under Langevin thermostat and two or four equilibrated configurations of ring polymers are added to the relaxed perovskite nanocrystal. With this initial configuration, simulations are done in three steps with timestep 1.0\,fs at 50K. First, we run 100\,ps in NVT for equilibration purpose with the lattice frozen and store configuration every 20\,ps. Next, starting from each stored configuration, lattice minimization is performed with the ring polymers frozen, which includes the effects of the lattice distortion with given configuration of ring polymers. Lastly, we run additional 500\,ps in NVT with the distorted lattice frozen. Simulations and minimizations using conjugate gradient method are performed using the LAMMPS package \cite{lammps}.


\section{Simulation details with static and dynamic screenings}

Under static and dynamic lattice screenings, simulations are performed with isolated electron and hole ring polymers. For the simulations under dynamic lattice screening, the Hamiltonian consists of two pieces, one from the electronic part of Hamiltonian given by Eq.\ref{biex_dHel} and the other which incorporates effects from harmonic phonons as written below 
\begin{equation}
\mathcal{H}_{\mathrm{eff}}
= \sum_{i,j} \mathcal{H}_{\mathrm{eff}}^{ij}
= - \sum_{i,j} \sum_{t \ne s} \sigma_{ij} \frac{\alpha_{ij} \beta \omega^2 \hbar^{5/2}}{n^2 \sqrt{8 m_{ij} \omega}} \frac{e^{-\beta\hbar\omega |t-s|/n}}{|\mathbf{x}_{i,t} - \mathbf{x}_{j,s}|}
\label{biex_dHeff}
\end{equation}
where $i,j \in \{e_1, h_1 \}$ or $i,j \in \{e_1, e_2, h_1, h_2 \}$ for exciton or biexciton, respectively, $t,s \in [1,n]$, n=300, and $\sigma_{ij}$ is either $+$1 or $-$1 if the charges of ring polymers $i$ and $j$ are the same or opposite with $\omega = 25.57 \mathrm{THz}$ \cite{sciadvzhu} and $m_{ij} = \sqrt{m_i m_j}$. The dimensionless Fröhlich coupling constant can be calculated using 
\begin{equation}
\alpha_{ij} = \frac{e^2}{\hbar} \sqrt{\frac{m_{ij}}{2 \hbar \omega}} \left( \frac{1}{\varepsilon_{\infty}} - \frac{1}{\varepsilon_s} \right) 
\end{equation}
where the resultant values are 2.65 and 2.76 for electrons and holes, respectively, with the static dielectric constant as $\varepsilon_s = 29.37$ \cite{sciadvzhu}. Detailed description and derivation of $\mathcal{H}_{\mathrm{eff}}$ can be found in Ref. \cite{parkjcp2022}.
For simulations with static lattice effects, the Hamiltonian is given by only Eq. \ref{biex_dHel}.

Simulations are performed using the LAMMPS package \cite{lammps} in an ensemble with constant number of atoms, volume, and temperature which is set to 50\,K using a Langevin thermostat with timestep 1.0\,fs and wall potentials are implemented as described above. 


\section{Details on binding energy calculations }

The exciton and biexciton binding energies can be computed from the average energy \cite{binde} of the system given by
\begin{equation}
\langle E \rangle = -\frac{\partial }{\partial \beta}
\ln \mathcal{Z} [\mathbf{x}_{eh}(\tau), \mathbf{x}_{\mathrm{lat}}(\tau)] 
\label{biex_avgE}
\end{equation}
where $\mathcal{Z}$ is the partition function of the system, $\mathbf{x}_{eh} = \{\mathbf{x}_{e_1}, \mathbf{x}_{h_1} \}$ or $\mathbf{x}_{eh} = \{\mathbf{x}_{e_1}, \mathbf{x}_{e_2}, \mathbf{x}_{h_1}, \mathbf{x}_{h_2} \}$ for exciton and biexciton, respectively. The derivative above produces two terms, the average kinetic energy $\langle E \rangle_{\mathrm{K}}$ and the average potential energy $\langle E \rangle_{\mathrm{P}}$, where to compute the kinetic energy, we use a virial estimator \cite{virial, parkjcp2022} to efficiently estimate the kinetic energy and avoid the large fluctuations arising from the subtraction of two diverging terms in path integral simulations. 
For the system with explicit perovskite nanocrystals, the resultant average potential and kinetic energies becomes
\begin{equation}
\langle E \rangle_{\mathrm{P}}
= \sum_{i\ne j} \langle \mathcal{H}_{\mathrm{C}}^{ij} \rangle
+ \langle U_{\mathrm{lat}} \rangle 
+ \sum_{i} \langle \mathcal{H}_{\mathrm{int}}^i \rangle 
\label{biex_avgpe1}
\end{equation}
and
\begin{equation}
\begin{aligned}
\langle E \rangle_{\mathrm{K}} 
=  \frac{3 }{2} k_{\mathrm{B}}T (N_{\mathrm{RP}} + N) 
+ \sum_{i} \sum_{t=1}^n  \left \langle \frac{(\mathbf{x}_{i,t} - \bar {\mathbf{x}}_{i})}{2} 
\bigg( \frac{1}{n} \frac{\partial 
\mathcal{H}_{\mathrm{int}}^i (\mathbf{x}_{i,t})}{\partial \mathbf{x}_{i,t}} + \sum_{j \ne i} \frac{\partial 
\mathcal{H}_{\mathrm{C}}^{i j}}{\partial \mathbf{x}_{i,t}} \bigg) \right \rangle
\label{biex_avgke1}
\end{aligned}
\end{equation}
where the summation index $i$ and $j$ run over all the charges, the number of charges $N_{\mathrm{RP}}$ is taken as either 2 or 4 for an exciton or biexciton, respectively, and $\bar {\mathrm{x}}$ represents the center of mass of the ring polymer. 

For the calculations with isolated ring polymers under dynamic screening, instead of the terms associated with the explicit lattice, the terms of effective interaction denoted as $\mathcal{H}_{\mathrm{eff}}^{ij}$ are involved in the formula. Since the Hamiltonian itself is a function of $\beta$, the derivative in Eq.\ref{biex_avgE} produces additional factors in the potential average energy
\begin{equation}
\langle E \rangle_{\mathrm{P}}
= \sum_{i\ne j} \langle \mathcal{H}_{\mathrm{C}}^{ij} \rangle
+ \sum_{ij} \left[ 2 \langle \mathcal{H}_{\mathrm{eff}}^{ij} \rangle + \langle \mathcal{H}_{\mathrm{eff}}^{ij} \rangle ' \right]
\label{biex_avgrppe2}
\end{equation}
where $\langle \mathcal{H}_{\mathrm{eff}}^{ij} \rangle '$ is defined as $\langle \mathcal{H}_{\mathrm{eff}}^{ij} \rangle $ multiplied by an additional factor of $\beta \hbar \omega |t-s|/n$ in the summation for each bead index $t$ and $s$ \cite{parkjcp2022}. 
The corresponding average kinetic energy is written as 
\begin{equation}
\begin{aligned}
\langle E \rangle_{\mathrm{K}}
=  \frac{3 }{2} k_{\mathrm{B}}T N_{\mathrm{RP}} + \frac{1}{2}  \sum_{i } \sum_{t=1}^n \Bigg \langle (\mathbf{x}_{i,t} - \bar {\mathbf{x}}_{i})  
 \times \bigg( \frac{\partial \mathcal{H}_{\mathrm{eff}}^{ii}}{\partial \mathbf{x}_{i,t}} + \sum_{j\ne i} \frac{ \partial \big( \mathcal{H}_{\mathrm{C}}^{i j} + 2\mathcal{H}_{\mathrm{eff}}^{ij}\big) }{\partial \mathbf{x}_{i,t}} \bigg) \Bigg \rangle 
\label{biex_avgrpke2}
\end{aligned}
\end{equation}
With static screening, the binding energies are determined by only $\mathcal{H}_{\mathrm{C}}$ term in Eq.\ref{biex_avgrppe2} and Eq.\ref{biex_avgrpke2}. 


In order to efficiently extract the exciton and biexciton binding energies with dynamic lattice screening, the perturbative approach is used on the ensembles obtained from the simulations under static lattice effect. 
In the path integral framework, with $\mathcal{Z}_{\mathrm{D}}$ and $\mathcal{Z}_{\mathrm{S}}$ as the partition functions characterized by the Hamiltonian defined as $\mathcal{H}_{\mathrm{el}} + \mathcal{H}_{\mathrm{eff}}$ and $\mathcal{H}_{\mathrm{el}}$ using Eq.\,\ref{biex_dHel} and \ref{biex_dHeff}, respectively,
the average energy under dynamic lattice screening is written as 
\begin{equation}
\langle E \rangle_{\mathrm{D}}
= \int \mathcal{D}[\mathbf{x}_{eh}] \, E \,\frac{e^{-\beta (\mathcal{H}_{\mathrm{el}} + \mathcal{H}_{\mathrm{eff}})}}{\mathcal{Z}_{\mathrm{D}}}
\end{equation}
where $\mathbf{x}_{eh}$ can be $\{\mathbf{x}_{e_1}, \mathbf{x}_{h_1} \}$ or $\{\mathbf{x}_{e_1}, \mathbf{x}_{e_2}, \mathbf{x}_{h_1}, \mathbf{x}_{h_2} \}$ for the system of exciton and biexciton, respectively. Using the rearrangement shown below, 
\begin{equation}
\begin{aligned}
\langle E \rangle_{\mathrm{D}}
 = \int \mathcal{D}[\mathbf{x}_{eh}] \, E \,\frac{e^{-\beta (\mathcal{H}_{\mathrm{el}} + \mathcal{H}_{\mathrm{eff}})}}{\mathcal{Z}_{\mathrm{D}}} 
\,\frac{\mathcal{Z}_{\mathrm{S}}}{\mathcal{Z}_{\mathrm{S}}} \, e^{\beta \mathcal{H}_{\mathrm{eff}} } \, e^{-\beta \mathcal{H}_{\mathrm{eff}} } 
 = \int \mathcal{D}[\mathbf{x}_{eh}] \, E \, \frac{e^{-\beta \mathcal{H}_{\mathrm{el}}}}{\mathcal{Z}_{\mathrm{S}}} \, \frac{\mathcal{Z}_{\mathrm{S}}}{\mathcal{Z}_{\mathrm{D}}} \, e^{-\beta \mathcal{H}_{\mathrm{eff}} }
\end{aligned}
\end{equation}
the average energy under dynamic lattice effect can be expressed in terms of the quantity averaged under static lattice effect as 
\begin{equation}
\langle E \rangle_{\mathrm{D}} 
= \frac{\mathcal{Z}_{\mathrm{S}}}{\mathcal{Z}_{\mathrm{D}}} \langle E \, e^{-\beta \mathcal{H}_{\mathrm{eff}} } \rangle_{\mathrm{S}}
\label{dynener}
\end{equation}
where $\langle \dots \rangle_{\mathrm{S}}$ indicates the expectation value taken within an ensemble under static lattice screening. Since $\mathcal{Z}_{\mathrm{D}}$ can be written in terms of $\mathcal{Z}_{\mathrm{S}}$ as
\begin{equation}
\begin{aligned}
\mathcal{Z}_{\mathrm{D}}
 = \int \mathcal{D}[\mathbf{x}_{eh}] \, e^{-\beta (\mathcal{H}_{\mathrm{el}} + \mathcal{H}_{\mathrm{eff}})}
\, \frac{e^{\beta \mathcal{H}_{\mathrm{eff}} } }{e^{\beta \mathcal{H}_{\mathrm{eff}} } }
\, \frac{\mathcal{Z}_{\mathrm{S}}}{\mathcal{Z}_{\mathrm{S}}} 
 = \mathcal{Z}_{\mathrm{S}} \int \mathcal{D}[\mathbf{x}_{eh}] \, \frac{e^{-\beta \mathcal{H}_{\mathrm{el}}}}{\mathcal{Z}_{\mathrm{S}}} \, e^{-\beta \mathcal{H}_{\mathrm{eff}} }
= \mathcal{Z}_{\mathrm{S}} \langle e^{-\beta \mathcal{H}_{\mathrm{eff}} } \rangle_{\mathrm{S}}
\end{aligned}
\end{equation}
the average energy given by Eq.\,\ref{dynener} becomes
\begin{equation}
\begin{aligned}
\langle E \rangle_{\mathrm{D}}
= \frac{\langle E \, e^{-\beta \mathcal{H}_{\mathrm{eff}} } \rangle_{\mathrm{S}} }{\langle e^{-\beta \mathcal{H}_{\mathrm{eff}}} \rangle_{\mathrm{S}}} 
 = \left[ \langle E \rangle_{\mathrm{S}} - \beta  \langle E \,\mathcal{H}_{\mathrm{eff}} \rangle_{\mathrm{S}} + \dots \right] 
\left[1 + \beta \langle \mathcal{H}_{\mathrm{eff}} \rangle_{\mathrm{S}} + \dots \right]
\label{biex_taylor}
\end{aligned}
\end{equation}
where the second equality is held by using Taylor expansion on exponential terms in the numerator and the denominator. 
Up to the first order, Eq.\ref{biex_taylor} becomes 
\begin{equation}
\langle E \rangle_{\mathrm{D}}
\approx  \langle E \rangle_{\mathrm{S}} - \beta \left[ \langle E \,\mathcal{H}_{\mathrm{eff}} \rangle_{\mathrm{S}} - \langle E \rangle_{\mathrm{S}} \, \langle E \,\mathcal{H}_{\mathrm{eff}} \rangle_{\mathrm{S}} \right] 
\end{equation}
with a compact expression as 
\begin{equation}
\langle E \rangle_{\mathrm{D}}
= \langle E \rangle_{\mathrm{S}} - \beta \langle \delta E \delta \mathcal{H}_{\mathrm{eff}} \rangle_{\mathrm{S}}
\end{equation}
which is the equation used to compute the average energy for exciton and biexciton binding energies under dynamic lattice screening. 

\vspace{5mm}
\bibliography{SI}

\newpage

\renewcommand{\arraystretch}{1.5}
\begin{table}
\centering
\begin{tabular}{ |c|c|c|c| }
\hline 
$i$ & \ $\varepsilon_{i}$ (kcal/mol) \ & \ $\sigma_{i} $ ($\mathrm{\AA}$) \ & \ $q_i$ ($e$) \ \\
\hline
\ Cs \ & 13.3381 & 2.927 & 0.86 \\
\ Pb \ & 0.2470 & 2.524 & 1.03 \\
\ Br \ & 0.2359 & 4.129 & $-$0.63 \\
\ Cs$_{\mathrm{surf}}$ \ & 13.3381 & 2.927 & $q_{\mathrm{surf}}$  \\
\hline
\end{tabular}
\caption{Lennard-Jones parameters for CsPbBr$_3$ perovskite nanocrystals. For the parameters not listed here can be calculated using $\varepsilon_{ij} = \sqrt{\varepsilon_{i} \varepsilon_{j}}$ and $\sigma_{ij} = (\sigma_i + \sigma_j)/2$ and the charge for surface Cs atoms is given by Eq.\ref{biex_qsur} for each size of nanocrystal.}
\label{biex_LJ}
\end{table}

\renewcommand{\arraystretch}{1.5}
\begin{table}
\centering
\begin{tabular}{ |c|c|c|c|c| }
\hline 
 & Cs & Pb & Br & Cs$^{\mathrm{surf}}$ \\
\hline
\ $e_1$, $e_2$ \ & \ 1.9350 \ & \ 1.0403 \ & \ 1.0403 \ & \ 1.9350 \ \\
\ $h_1$, $h_2$ \ & \ 1.0403 \ & \ 1.0403 \ & \ 1.1655 \ & \ 1.0403 \ \\
\hline
\end{tabular}
\caption{Pseudopotential cutoff parameters $r_{\mathrm{cut}}$ in units of $\mathrm{\AA}$ for the interaction between quasiparticles (two electrons and two holes) with each type of atoms in the lattice.} 
\label{biex_pseudo}
\end{table}

